\documentclass[aps,twocolumn,float,prd,psfig,amsmath,amssymb]{revtex4-2}

\usepackage[dvipdfmx]{graphicx}
\usepackage{dcolumn}
\usepackage{bm}
\usepackage{braket}
\usepackage{breqn}
\usepackage{color}

\newcommand{\beq}{\begin{equation}}
\newcommand{\beqa}{\begin{eqnarray}}
		  \newcommand{\eeq}{\end{equation}}
\newcommand{\eeqa}{\end{eqnarray}}
\newcommand{\lla}{\left\langle}

\newcommand{\rra}{\right\rangle}

\begin{document}


\title{Measuring the maximally allowed polarization states of the isotropic stochastic gravitational wave background with the ground-based detectors}

\author{Hidetoshi Omiya}
\email{omiya@tap.scphys.kyoto-u.ac.jp}
\affiliation{Department of Physics$,$ Kyoto University$,$ Kyoto 606-8502$,$ Japan}
\author{Naoki Seto}
\email{seto@tap.scphys.kyoto-u.ac.jp}
\affiliation{Department of Physics$,$ Kyoto University$,$ Kyoto 606-8502$,$ Japan}

\date{\today}

\begin{abstract}
 We discuss  the polarizational study of isotropic  gravitational wave backgrounds with the second generation detector network, paying special attention to the impacts of adding LIGO-India.    The backgrounds can be characterized by at most five spectral components (three parity-even ones and two parity-odd ones).  They can be algebraically decomposed  through the difference of the corresponding overlap reduction functions defined for the individual spectra.  We newly identify two interesting relations between the overlap reduction functions, and these relations generally  hamper the algebraic decomposition in the low frequency regime $f\lesssim 30$Hz. We also find that LIGO-India can significantly   improve the network sensitives to the odd spectral components.  
\end{abstract}


\maketitle


\section{Introduction}

A stochastic gravitational wave background is one of the primary targets of gravitational wave detectors. There exist a large number of theoretical predictions for generation processes such as an inflationary expansion \cite{Starobinsky:1979ty,PhysRevLett.99.221301, Kuroyanagi:2008ye, Cook:2011hg}, a phase transition \cite{Kamionkowski:1993fg,Caprini:2007xq}, and distant unresolved binaries \cite{Farmer:2003pa,Zhu:2012xw} (for other sources, see \cite{Christensen:2018iqi,Kuroyanagi:2018csn}). Many of these backgrounds were generated in strong gravity regimes or high energy states and could be a good probe for physics in an extreme environment. Note also that these backgrounds are expected to be highly isotropic.

In General Relativity (GR), we only have the two tensor degrees of freedom, the + and $\times$ modes. In contrast, some alternative theories of gravity predict additional polarization modes; the two vector ($x$ and $y$) modes and the two scalar ($b$ and $l$) modes \cite{Berti:2015itd}.   Therefore, through a polarization study of background, we might detect a signature of modification to GR  \cite{Corda:2008si,Callister:2017ocg} (see \cite{Chatziioannou:2012rf, Isi:2017fbj,Takeda:2019gwk,Hagihara:2019ihn,Takeda:2020tjj, Takeda:2021hgo} for studies on the polarization of gravitational waves from compact binary). Furthermore, even if GR is not modified at present,  a parity violation process in the early universe could generate an asymmetry between right- and left-handed polarization patterns \cite{Lue:1998mq,Alexander:2004us,Kahniashvili:2005qi,Satoh:2007gn,Adshead:2012kp,Dimastrogiovanni:2016fuu,Ellis:2020uid,Okano:2020uyr}. 

The cross-correlation analysis is an efficient method for detecting a weak gravitational wave background \cite{Flanagan:1993ix, Allen:1997ad,Romano:2016dpx}. By taking products of data streams of noise-independent detector pairs, we can gradually improve the sensitivity to a background by increasing observational time.   When the gravitational wave frequency is much longer than the arm lengths of detectors \cite{Omiya:2021zif}, the two scalar modes are observationally non-separable, and we can generally measure the five background spectra $I_{T},I_{V},I_{S},W_{T},$ and $W_{V}$. The three spectra $I_{T},I_{V},$ and $I_S$ represent the total intensity of the tensor, vector, and scalar modes. The remaining two spectra $W_{T}$ and $W_{V}$ correspond to the Stokes ``V" parameters which probe the degrees of circular polarization of the tensor and vector modes. In this paper, we utilize $W$ for the ``V" parameter to avoid confusion with the vector modes. Since the spectra $I_{T},I_{V},$ and $I_S$ transform as parity even quantities and the $W_{T}$ and $W_V$ transform as parity odd quantities, we refer to the former three as parity even spectra and the later two as parity odd spectra.  

At the correlation analysis, we can measure linear combinations of the five spectra with the five coefficients known as the overlap reduction functions (ORFs). The ORFs characterize the sensitivities to the corresponding spectra and depend on gravitational wave frequency as well as the relative configuration of the two pairwise detectors.   
We apply the parity even/odd classification also to the five ORFs.

For probing the existence of the anomalous polarization spectra $I_V$, $I_S$, $W_T$, and $W_V$,  we desire to clean the contribution from the standard spectrum $I_T$ (see also \cite{KAGRA:2021kbb}  for a maximum likelihood analysis). In addition, we prefer to break down the four anomalous modes and measure them separately. Our strategy in this paper is to utilize the difference between the five ORFs and algebraically decompose the five spectra by taking appropriate linear combinations of the correlation products from multiple pairs (originally proposed in \cite{Seto:2006dz}). 
We mainly study the prospects of this algebraic scheme with the second generation detector network. We pay special attention to the impacts of adding LIGO-India as the fifth detector. 

In the middle of our study, we newly identify two degenerate relations between the ORFs. The first one is for the three even ORFs, and the second one is for the two odd ORFs. These two relations generally limit the performance of the algebraic decomposition in the low frequency regime $f\lesssim 30$Hz. 
On the other hand, LIGO-India can largely mitigate the damage associated with the degeneracy for the even ORFs,  because of its relatively remote location from the two other LIGO detectors. Furthermore, LIGO-India can significantly improve the sensitivities to the odd spectra. 

This paper is organized as follows. In Sec.~\ref{sec:2}, we review the polarization decomposition of an isotropic background and present the analytical expressions for the associated ORFs. We also explain our two new findings with respect to the ORFs.
In Sec.~\ref{sec:4}, we concretely study the geometry of the second generation terrestrial detector network,  including LIGO-India. In Sec.~\ref{sec:4A}, we review the correlation analysis, primarily focusing on the evaluation of the signal-to-noise ratio. In Sec.~\ref{sec:4B}, we explain the algebraic decomposition scheme for multiple spectral components. In Secs.~\ref{sec:7} and ~\ref{sec:8}, we apply this scheme to the second generation ground-based detector network. We discuss how the sensitivity depends on the target polarization spectra and the network combinations. Finally, in Sec.~\ref{sec:9}, we summarize our paper.

\section{Basic Quantities}\label{sec:2}

Following our preceding work  \cite{Omiya:2021zif} on formal aspects,  we first review the basic ingredients for the correlated signals of stochastic backgrounds with  ground-based detectors. Since our universe is highly isotropic and homogeneous, the monopole components of the backgrounds are assumed to be our primary target. In addition, because the observed speed of gravitational wave $v_g$ is close to the speed of light  $c$, we set $v_g = c$. 

In Sec.~\ref{sec:2A}, we describe the polarization states of the isotropic backgrounds and introduce the five relevant  spectra $I_T, I_V, I_S,W_T,$ and $W_V$. In Sec.~\ref{sec:2B}, we discuss the ORFs which characterize the correlated response of pairwise detectors to the backgrounds. In Sec.~\ref{sec:2C}, we give analytic expressions of the ORFs for the ground-based detectors. In Secs.~\ref{sec:2D} and~\ref{sec:2E}, we discuss their asymptotic behaviors. In   Secs.~\ref{sec:2-F} and~\ref{sec:2-G}, we report our two new findings on the ORFs. 

\subsection{Polarization states of a stochastic gravitational wave background}\label{sec:2A}

We start with the plane wave decomposition of the metric perturbation $h_{ij}$  generated by the gravitational waves
\begin{align}
\begin{aligned}
	h_{ij}(t,\bm{x}) = &\sum_{P} \int df \int d\bm{\Omega}\\
	& \times \tilde{h}_P(f,\bm{\Omega}) \bm{e}_{P,ij}(\bm{\Omega}) e^{-2\pi i f (t - \bm{\Omega} \cdot \bm{x}/c)}~,
\end{aligned}
\end{align}
where $\bm{\Omega}$ is the unit vector for the propagation direction, normalized by $\int d\bm{\Omega} = 4 \pi$. Here, $\bm{e}_P$ ($P=+,\times,x,y,b$ and $l$) represent the polarization tensors given by
\begin{align}\label{eq:3}
\begin{aligned}
	\bm{e}_{+} &= \bm{m} \otimes \bm{m} - \bm{n}\otimes \bm{n}~, & \bm{e}_{\times} &= \bm{m} \otimes \bm{n} + \bm{n}\otimes \bm{m}~,\\
	 \bm{e}_{x} &= \bm{\Omega} \otimes \bm{m} + \bm{m}\otimes \bm{\Omega}~,& \bm{e}_y &= \bm{\Omega} \otimes \bm{n} + \bm{n}\otimes \bm{\Omega}~,\\
	 \bm{e}_b &= \sqrt{3}(\bm{m} \otimes \bm{m} + \bm{n}\otimes \bm{n})~,& \bm{e}_l &= \sqrt{3}(\bm{\Omega}\otimes\bm{\Omega})
\end{aligned}
\end{align}
with the  orthonormal vectors $\bm m$ and $\bm n$ in addition to $\bm \Omega$ (see \cite{Will:1993ns} for geometrical interpretation of these modes). { Note that our definitions for $e_b$ and $e_l$ are different from the conventional one such as used in \cite{Callister:2017ocg} (see also Appendix in  \cite{Omiya:2020fvw}).} They are written by the standard polar coordinates  $(\theta,\phi)$ as
\begin{align}
	\bm{\Omega} &= \left(
	\begin{array}{c}
	\sin\theta \cos\phi\\
	\sin\theta\sin\phi\\
	\cos\theta
	\end{array}
	\right)~,\\
	\bm{m} &= \left(
	\begin{array}{c}
	\cos\theta \cos\phi\\
	\cos\theta\sin\phi\\
	-\sin\theta
	\end{array}
	\right)~,\\
	\bm{n} &= \left(
	\begin{array}{c}
	-\sin\phi\\
	\cos\phi\\
	0
	\end{array}
	\right)~.
\end{align}
In Eq. (\ref{eq:3}),  the labels $P=+,\times$ correspond to the tensor ($T$) modes,  $P = x,y$ to the vector ($V$) modes,  and $P = b,l$  to the scalar ($S$) modes. Note that GR predicts only the tensor modes.    However, numerous alternative theories of gravity allow the existence of the remaining  $V$ and $S$ modes.

For a stochastic background, the expansion coefficients $\tilde{h}_P$ can be regarded as random quantities. Their statistical properties are  specified by the power spectrum matrix $\braket{\tilde{h}_P(f,\bm{\Omega})\tilde{h}_{P'}^*(f',\bm{\Omega})}$ with no correlation between $T,V$ and $S$ modes  for statistically isotropic backgrounds \cite{Omiya:2021zif}.  In the case of the tensor modes ($P,P' = +,\times$), the matrix can be written in terms of  the Stokes parameters as \cite{Seto:2008sr} 
\begin{align}\label{eq:PST}
\begin{aligned}
	\braket{\tilde{h}_{P}(f,\bm{\Omega})\tilde{h}_{P'}^*(f',\bm{\Omega'})} =& \frac{1}{2}\delta_{\Omega\Omega'}\delta(f-f')\\
	&\times \left(
	\begin{array}{cc}
		I_T+Q_T & U_T-iW_T\\
		U_T+iW_T & I_T-Q_T
	\end{array}
	\right)_{PP'}~.
\end{aligned}
\end{align}
In   the standard literature of polarization, the chiral asymmetry is usually denoted as the Stokes ``$V$'' parameter. In  this paper, we apply  the notation $V$ to represent the vector modes, and use $W$ for the chiral asymmetry. Note that the combinations $Q_T\pm iU_T$ do not  have   isotropic components, as understood from their transformation properties  \cite{Seto:2008sr, Omiya:2021zif}. We  thus drop them hereafter. 

In Eq.  (\ref{eq:PST}), we  use the coefficients $\tilde{h}_{P}(f,\bm{\Omega})$  for the linear polarization bases $(\bm{e}_{+},\bm{e}_{\times})$. 
However, the physical meaning of the $W$ parameter becomes transparent 
 by introducing the circular (right- and left-handed) polarization  bases given by 
\begin{align}\label{eq:7}
	\bm{e}_R^T &= \frac{1}{\sqrt{2}}\left(\bm{e}_+ + i \bm{e}_\times \right)~, & \bm{e}_L^T &= \frac{1}{\sqrt{2}}\left(\bm{e}_+ - i \bm{e}_\times \right)~,
\end{align}
with the corresponding coefficients
\begin{align}\label{eq:8}
	\tilde{h}^{T}_R(f,\bm{\Omega}) &= \frac{1}{\sqrt{2}}\left(\tilde{h}_{+}(f,\bm{\Omega}) - i \tilde{h}_{\times}(f,\bm{\Omega})\right)~,\\
	\tilde{h}^{T}_L (f,\bm{\Omega})&= \frac{1}{\sqrt{2}}\left(\tilde{h}_{+}(f,\bm{\Omega})+ i \tilde{h}_{\times}(f,\bm{\Omega})\right).\label{eq:9}
\end{align}
We then  have
\begin{align}
	I_{T} &= \braket{\tilde{h}^{T}_R\tilde{h}_R^{T*}} + \braket{\tilde{h}^{T}_L \tilde{h}_L^{T*}}~, \\
	W_{T} &= \braket{\tilde{h}^{T}_R \tilde{h}_R^{T*}} - \braket{\tilde{h}^{T}_L \tilde{h}_L^{T*}},
\end{align}
omitting apparent delta functions.  
These expressions show that the spectra  $I_{T}$ and $W_{T}$ characterize the total and asymmetry of the amplitudes of the right- and left-handed polarization patterns of the tensor  modes. Since the parity transformation interchanges the right-and left-handed waves, we resultantly have  $I_T'=I_T$ and $W_T'=-W_T$ ($'$ representing  parity transformed quantities).

For the vector modes, we can repeat almost the same arguments as Eqs.~\eqref{eq:PST}-\eqref{eq:9} and obtain
\begin{align}
	I_{V} &= \braket{\tilde{h}^{V}_R\tilde{h}_R^{V*}} + \braket{\tilde{h}^{V}_L \tilde{h}_L^{V*}}~, \\
	W_{V} &= \braket{\tilde{h}^{V}_R \tilde{h}_R^{V*}} - \braket{\tilde{h}^{V}_L \tilde{h}_L^{V*}}
\end{align}
with the correspondences $I_V'=I_V$ and $W_V'=-W_V$  for  the parity transformation.

For the scalar modes ($P,P' = b,l$), considering their potential correlation,  we  can generally put 
\begin{align}
\begin{aligned}
	\braket{\tilde{h}_{P}(f,\bm{\Omega})\tilde{h}_{P'}^*(f',\bm{\Omega'})} =& \frac{1}{2}\delta_{\Omega\Omega'}\delta(f-f')\\
	&\times \left(
	\begin{array}{cc}
		I_b & C_S\\
		C_S^* & I_l
	\end{array}
	\right)_{PP'}~.
\end{aligned}
\end{align}
described  by the four real parameters in the power spectra. In reality, as long as the low  frequency approximation  is valid ($f\ll (2\pi L/c)^{-1}$, $L$: the arm length),  only  the combination
\begin{align}
	I_S \equiv \frac{1}{2}(I_b + I_l - C_S - C_S^*)~,
\end{align}
appears in the correlation analysis \cite{Omiya:2021zif}.   Therefore, in the following, we  keep only $I_S$ for the scalar modes. Because of its spin-0 nature, we  also have $I_S'=I_S$  for the parity transformation.

Up to now, we see that an isotropic  background is characterized by the five quantities $I_T,I_V,I_S,W_T,$ and $W_V$. 
Here, we introduce another commonly used representation for the magnitudes of these spectra. In GR, the amplitude $I_T(f)$ can be simply related  to the energy density of the background. More specifically, 
 with the Hubble parameter $H_0$,  we have the relation
\begin{align}\label{omega}
	\Omega_{GW}^{I_T}(f) = \left(\frac{32\pi^3}{3 H_0^2}\right) f^3 I_T(f)~
\end{align}
for the energy density of the  background per logarithmic frequency (normalized by critical density of universe)~\cite{Flanagan:1993ix, Allen:1997ad}.
 In  a modified theory of gravity,  the relation~\eqref{omega} for the energy density might be invalid   \cite{Isi:2018miq}.   However, we are not directly interested in the actual energy density of the backgrounds, and thus continue to use Eq.~\eqref{omega}  as the definition of $\Omega_{GW}^{I_T}(f) $. 
 Similarly,  we use the effective energy densities 
\begin{align}
	\Omega_{GW}^{I_V}(f) &\equiv \left(\frac{32\pi^3}{3 H_0^2}\right) f^3 I_V(f)~,\label{e17}\\
	 \Omega_{GW}^{I_S}(f) &\equiv \left(\frac{32\pi^3}{3 H_0^2}\right) f^3 I_S(f)~,\\
	\Omega_{GW}^{W_T}(f) &\equiv \left(\frac{32\pi^3}{3 H_0^2}\right) f^3 W_T(f)~,\\
	 \Omega_{GW}^{W_V}(f) &\equiv \left(\frac{32\pi^3}{3 H_0^2}\right) f^3 W_V(f)~.\label{omegaWV}
\end{align}
If the left handed modes dominate  the right handed ones, we  have $\Omega_{GW}^{W_T}(f) <0$ (and  $\Omega_{GW}^{W_V}(f) <0$). 

\subsection{Correlation Analysis}\label{sec:2B}

Now we discuss how to detect the five spectral components  by using multiple  interferometers. In the low frequency regime ($f\ll (2\pi L /c)^{-1}$), the response of an interferometer $A$ (at the position $\bm{x}_A$) can be modeled as~\cite{Omiya:2021zif}
\begin{align}\label{eq:GWsig}
	h_A(f) = \bm{D}_{A}^{ij} \tilde{h}_{ij}(f, \bm{x}_A)~,
\end{align}
with the beam pattern function
\begin{align}
	\bm{D}_{A} = \frac{\bm{u}_A \otimes \bm{u}_A - \bm{v}_A\otimes \bm{v}_A}{2}~.
\end{align}
Here, $\tilde{h}_{ij}(f, \bm{x}_A)$ is the metric perturbation of the background at the detector, and the two unit vectors $\bm{u}_A$ and $\bm{v}_A$ represent the two arm directions of the detector. 

By correlating data streams of multiple detectors, we can statistically amplify the  background signals relative to  the detector noises (closely discussed in Sec. IV). We denote the correlation product of two detector $A$ and $B$ by
\begin{align}\label{eq:GWcorr}
	C_{AB}(f) \equiv \braket{h_A(f) h_B^*(f)}~
\end{align}
(again omitting delta functions).
Leaving only the monopole components of the background, we obtain
\begin{align}\label{eq:GWexp}
\begin{aligned}
	C_{AB}(f) &= \frac{4\pi}{5} \left(\sum_{P=T,V,S}\gamma^{I_P} I_P + \sum_{P=T,V}\gamma^{W_P} W_P\right)~.
\end{aligned}
\end{align}
Here, $\gamma^{I_P}$ and $\gamma^{W_P}$ are the ORFs which characterize the correlated response of two detectors to the relevant   components of an isotropic background. They are written as
\begin{align}\label{25}
	\gamma^{I_P}_{AB}&(f) \equiv \bm{D}_{A,ij}\bm{D}_{B,kl} \Gamma^{I_P}_{ijkl}~,\\
	\gamma^{W_P}_{AB}&(f) \equiv \bm{D}_{A,ij}\bm{D}_{B,kl} \Gamma^{W_P}_{ijkl}~,
\end{align}
with  the angular integrals
\begin{align}
	\Gamma^{I_T}_{ijkl} &=\frac{5}{8 \pi} \int d\bm{\Omega} (e_{+,ij} e_{+,kl} + e_{\times,ij} e_{\times,kl}) e^{i y \bm{\Omega}\cdot \hat{\bm{d}}}~,\label{27}\\
	\Gamma^{I_V}_{ijkl} &=\frac{5}{8 \pi} \int d\bm{\Omega} (e_{x,ij} e_{x,kl} + e_{y,ij} e_{y,kl}) e^{i y \bm{\Omega}\cdot \hat{\bm{d}}}~,\\
	\Gamma^{I_S}_{ijkl} &=\frac{5}{8 \pi} \int d\bm{\Omega} (e_{b,ij} e_{b,kl} + e_{l,ij} e_{l,kl}) e^{i y \bm{\Omega}\cdot \hat{\bm{d}}}~,\\
	\Gamma^{W_T}_{ijkl} &=-\frac{5 i}{8 \pi} \int d\bm{\Omega} (e_{+,ij} e_{\times,kl} - e_{\times,ij} e_{+,kl}) e^{i y \bm{\Omega}\cdot \hat{\bm{d}}}~,\\
	\Gamma^{W_V}_{ijkl} &=-\frac{5 i}{8 \pi} \int d\bm{\Omega} (e_{x,ij} e_{y,kl} - e_{y,ij} e_{x,kl}) e^{i y \bm{\Omega}\cdot \hat{\bm{d}}}~. \label{31}
\end{align}
Here, we put $d\equiv |{\bm x}_A-{\bm x}_B|$,  $\hat{\bm{d}} \equiv  ({\bm x}_A-{\bm x}_B)/d$ and $y = 2\pi f d/c$. 

As already in Eq.~\eqref{eq:GWexp}, we will use the label $P$ for the polarization modes $P=(T,  V, S)$, extending it from the original patterns $P=(+,\times,x,y,b,l)$.  
In  addition, we introduce the label $Q$ to represent all the five spectral modes $(I_T,I_V,I_S,W_T,W_V)$ of interest.
For notational simplicity,  we also omit the labels for the detectors in obvious cases. 

\subsection{ORFs for ground-based detectors}\label{sec:2C}

\begin{figure}
\centering
\includegraphics[keepaspectratio, scale=0.2]{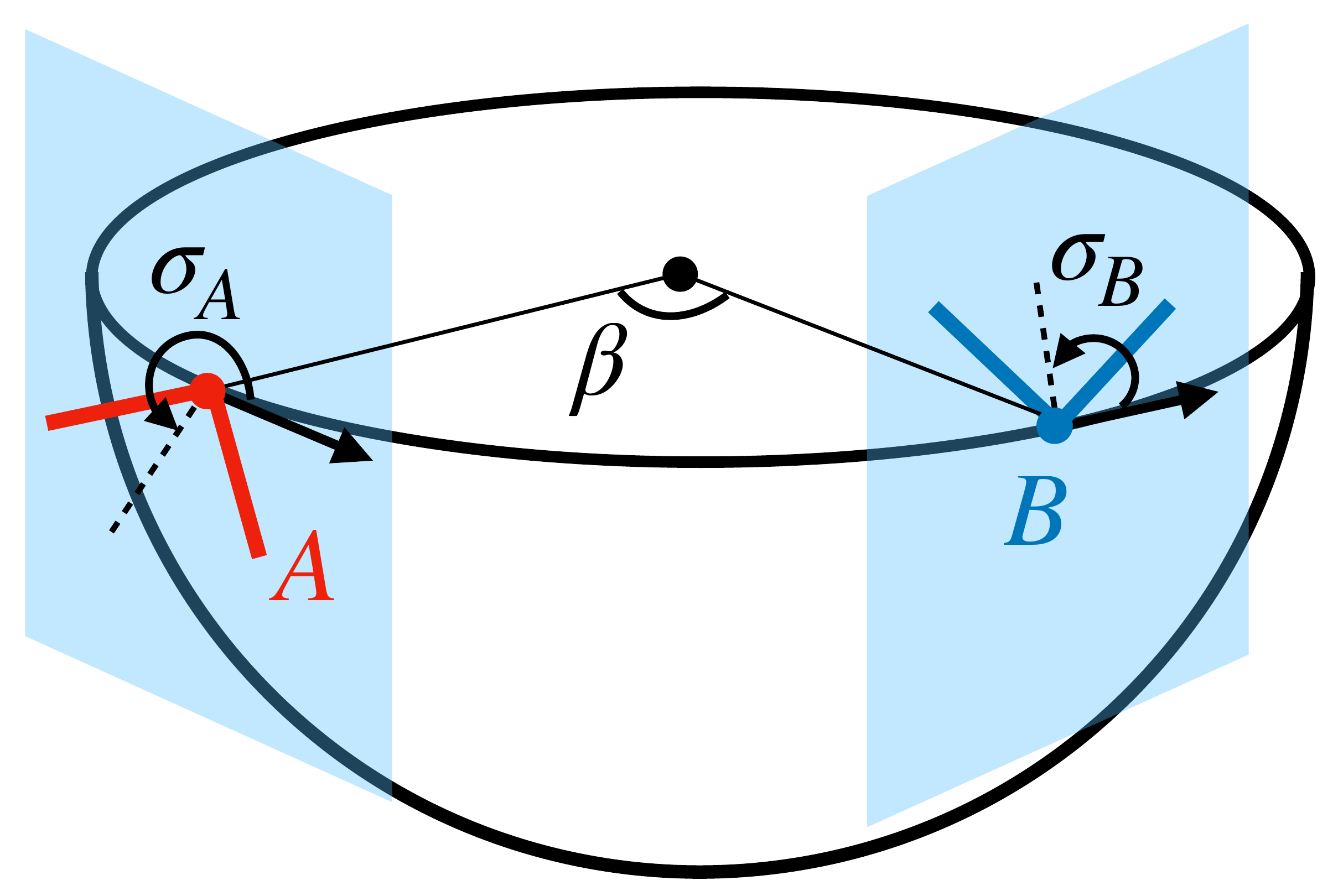}
\caption{The relative geometry of the ground-based detector pair $A$ and $B$. The two detectors are on the same great circle and their detector planes are tangential to the earth sphere. The opening angle $\beta$ is measured from the center of the Earth. The angles $\sigma_A$ and $\sigma_B$ correspond to the orientations of the bisectors of the two arms (dotted line) measured counter clock wisely relative to the great circle.}    
\label{fig:groundangle}
\end{figure}

Now we focus on the ground-based detectors that are assumed to be tangential to the Earth sphere of the radius $R_E=6400$km.   As shown in Fig. 1,  the relative geometry of two interferometers $A$ and $B$ are fully characterized by three angles $\beta$, $\sigma_A$ and $\sigma_B$ (following the convention in \cite{Flanagan:1993ix}).   The angle $\beta$ represents the opening angle between the two detectors, measured from the center of the Earth, and we have $d=2R_E\sin(\beta/2)$. Meanwhile, the angle $\sigma_{A}$  shows the orientation of the bisector of the two arms of  the detector $A$ measured counterclockwise relative to the great circle joining the two detectors. The angle $\sigma_B$ is defined similarly. Below, instead of $\sigma_A$ and $\sigma_B$, we use the angles $\Delta$ and $\delta$
\begin{align}
	\Delta &\equiv \frac{\sigma_A + \sigma_B}{2}~, & \delta &\equiv \frac{\sigma_A - \sigma_B}{2}~,
\end{align}
following the  standard convention. 

The close expressions of the ORFs are presented in    \cite{Omiya:2021zif} as 
\begin{align}\label{eq:B6}
	\gamma^{I_P}_{} &= \Theta_\Delta^{P}(y,\beta) \cos4 \Delta + \Theta_\delta^{P}(y,\beta)\cos4\delta~, & (P &= T,V,S)~,\\
	\gamma^{W_P}_{} &= \Xi^{P}(y,\beta) \sin4 \Delta~, & (P &= T,V)~.
\end{align}
Here the angles $\delta$ and $\Delta$  appear only in the forms $\cos 4\delta, \cos4\Delta,$ and $\sin4\Delta$, reflecting certain   symmetries \cite{Seto:2007tn}. The coefficients $\Xi^P, \Theta_\Delta^P,$ and $\Theta_\delta^P$ are given by
\begin{align}\label{eq:THDtensor}
	\Theta^T_\Delta(y,\beta) &= - \sin^4 \left(\frac{\beta}{2}\right) j_0(y)\cr
	& - \frac{5}{56}(-9 + 8\cos\beta + \cos2\beta) j_2(y)\cr
	& - \frac{1}{896}(169 + 108 \cos\beta + 3 \cos2\beta) j_4(y)~,\\
	\Theta^V_\Delta(y,\beta) &= - \sin^4 \left(\frac{\beta}{2}\right) j_0(y)\cr
	&+ \frac{5}{112}(-9 + 8 \cos\beta  + \cos2\beta) j_2(y)\cr
	&+ \frac{1}{224}(169 + 108 \cos\beta + 3 \cos2\beta) j_4(y)~,\\
	\Theta^S_\Delta(y,\beta) &= - \sin^4 \left(\frac{\beta}{2}\right) j_0(y) \cr
	&+ \frac{5}{56}(-9 + 8\cos\beta + \cos 2\beta) j_2(y) \cr
	&- \frac{3}{448}(169 + 108 \cos\beta + 3 \cos2\beta) j_4(y)~,
\end{align}
\begin{align}
	\Theta^T_{\delta}(y,\beta) &= \cos^4\left(\frac{\beta}{2}\right) \left(j_0(y) + \frac{5}{7}j_2(y) + \frac{3}{112} j_4 (y) \right)~,\\
	\Theta^V_{\delta}(y,\beta) &=\cos^4\left(\frac{\beta}{2}\right) \left(j_0(y) - \frac{5}{14}j_2(y) - \frac{3}{28} j_4 (y) \right)~,\\
	\Theta^S_{\delta}(y,\beta) &=\cos^4\left(\frac{\beta}{2}\right) \left(j_0(y) - \frac{5}{7}j_2(y) + \frac{9}{56} j_4 (y) \right)~,
\end{align}

\begin{align}
		\Xi^T(y,\beta) &= \sin\left(\frac{\beta}{2}\right)\left((1-\cos\beta)j_1(y) - \frac{7 + 3 \cos \beta}{8}j_3(y)\right)~,\\
		\label{eq:XIvector}
	\Xi^V(y,\beta) &= \frac{1}{2}\sin\left(\frac{\beta}{2}\right)\left((1-\cos\beta)j_1(y) + \frac{7 + 3 \cos \beta}{2}j_3(y)\right) 
\end{align}
with the spherical Bessel functions $j_n(y)$.

\subsection{Asymptotic Behaviors at $y\to \infty$}\label{sec:2D}
In this subsection, we briefly discuss the asymptotic profiles of the ORFs at  $y\to \infty$,  based on  Eqs.~\eqref{eq:B6}-\eqref{eq:XIvector}. 

For the spherical Bessel functions, at large $y$, we have the following  correspondences 
\begin{align}
\label{eq:asympbessel}
	j_{2 l}(y) &\propto \frac{\sin y }{y}~, &
	j_{2 l+1}(y) &\propto \frac{\cos y}{y}~, 
\end{align}
Then, we can put
\begin{align}\label{eq:asymp} 
	\gamma^{I_P} &\to  C_{I_{P}} \frac{\sin y}{y}~, &
	\gamma^{W_P} &\to C_{W_{P}} \frac{\cos y}{y}
\end{align}
with the coefficients $C_{I_P}$ and $C_{W_P}$ presented shortly. 
Roughly speaking, these relations show the phase offset of $\sim\pi/2$ (as  in the combination of  $\sin y$ and $\cos y$), 
depending on the two parity types of the background spectra  $I^P$ and $W^P$. 

We can readily evaluate the coefficients $C_{I_P}$ and $C_{W_P}$ as follows; 
\begin{align}
	C_{I_T} &= \frac{5}{128} \Bigl(8 \cos ^4\left(\frac{\beta
   }{2}\right) \cos 4 \delta \cr
   &- (\cos
   2 \beta - 28 \cos\beta +35) \cos 4 \Delta \Bigr)~,\\
   C_{I_V} &= \frac{5}{8} \cos
   ^2\left(\frac{\beta }{2}\right) \Bigl(2 \cos
   ^2\left(\frac{\beta }{2}\right) \cos 4 \delta \cr
   &-(\cos
   \beta -3) \cos 4 \Delta \Bigr)~,\\
   C_{I_S} &= \frac{15}{8} \cos
   ^4\left(\frac{\beta }{2}\right) (\cos 4 \delta -\cos
   4 \Delta )~,
\end{align}
\begin{align}
	C_{W_T} &= -\frac{5}{16} \left(-\sin \left(\frac{3 \beta
   }{2}\right)+7 \sin \left(\frac{\beta }{2}\right)\right)
   \sin 4 \Delta~,\label{56}\\
   C_{W_V} &= \frac{5}{2} \sin \left(\frac{\beta
   }{2}\right) \cos ^2\left(\frac{\beta }{2}\right) \sin 4
   \Delta~.\label{57}
\end{align}
We  have $ C_{W_T}\cdot  C_{W_V}\le 0$.
Notice that $C_{W_T}$ and $C_{W_V}$ vanish at $\beta = 0$. Two detectors on a plane are apparently mirror symmetric, and thus blind to the parity odd polarizations.

\subsection{Asymptotic Behaviors at $y\to 0$}\label{sec:2E}

At  the opposite limit, $y\to 0$,  we  have
\begin{align}
	\gamma^{I_{T,V,S}}(y) &\to 2D_{A,ij}D_B^{ij}\\
	&=-\sin^4\left(\frac{\beta}{2}\right)\cos 4\Delta + \cos^4\left(\frac{\beta}{2}\right)\cos 4\delta~, \label{ge0}\\
	\gamma^{W_T}(y) &\to 2\sin^3\left(\frac{\beta}{2}\right)y\sin4\Delta ~,\label{go1}\\
	\gamma^{W_V}(y) &\to \sin^3\left(\frac{\beta}{2}\right)y\sin4\Delta~.\label{go2}
\end{align}
The first expression shows the degeneracy of the  parity even ORFs. Meanwhile, the parity odd ORFs vanish at $y\to0$, due to the parity symmetry of a network at the same place  with $d=0$ \cite{Omiya:2021zif}. {Thus, a network becomes blind to the parity odd polarizations for small $y$. }

\subsection{ Trinity degeneracy of even ORFs at the sub-leading order $O(y^2)$}\label{sec:2-F}

At  the sub-leading order $O(y^2)$ (or  equivalently $O(f^2)$), we can easily confirm a cancellation for the three even ORFs and have 
\begin{equation}
	\gamma^{I_T}(y)-4\gamma^{I_V}(y) +3\gamma^{I_S}(y)= O(y^4)~.\label{deg}
\end{equation}
This trinity degeneracy will later play an important role in the spectral decomposition of the three even spectra.

\subsection{Degeneracy of odd ORFs at 13Hz}\label{sec:2-G}

For detectors on the Earth, we can put $y=\zeta\sin(\beta/2)$ with $\zeta\equiv 4\pi R_Ef/c$. In Fig. 2, we present a contour plot for the ratio between the odd ORFs 
\begin{equation}
\frac{\gamma_{W_T}}{\gamma_{W_V}}=\frac{\Xi^T(y,\beta)}{\Xi^V(y,\beta)}\equiv \Theta (\zeta,\beta).
\end{equation}
At the left end, we can see the  limit $\lim_{\zeta\to0}\Theta(\zeta,\beta)=2$ following from Eqs.~\eqref{go1} and \eqref{go2}. 

Surprisingly,  the function $\Theta$ depends very weakly on $\beta$ around $\zeta=3.57$, as shown with the almost vertical contour $\Theta=1.26$ in Fig. 2. Indeed, along this contour, the variation of $\zeta$ is within $\pm0.01$.
Later, we will find that the odd spectral decomposition practically collapses around $\zeta=3.57$, corresponding to $f=13$Hz for the Earth's radius $R_E=6400$km. This  anathematic frequency is intrinsic to ground-based detectors. 

In space, we might realize a detector network composed by multiple LISA-like units orbiting around the Sun~\cite{Audley:2017drz,Hu:2017mde} (see also~\cite{Luo:2015ght}). For their typical orbital configuration,  we need at least three separated units for fully decomposing the five polarization spectra, and these units contact with a virtual  sphere of radius 1.15 a.u.~\cite{Seto:2020zxw,Omiya:2020fvw,Liu:2022umx} (see also \cite{Seto:2020mfd}).  In this case, the anathematic frequency becomes 0.57mHz.  

\begin{figure}
\centering
\includegraphics[keepaspectratio, scale=0.5]{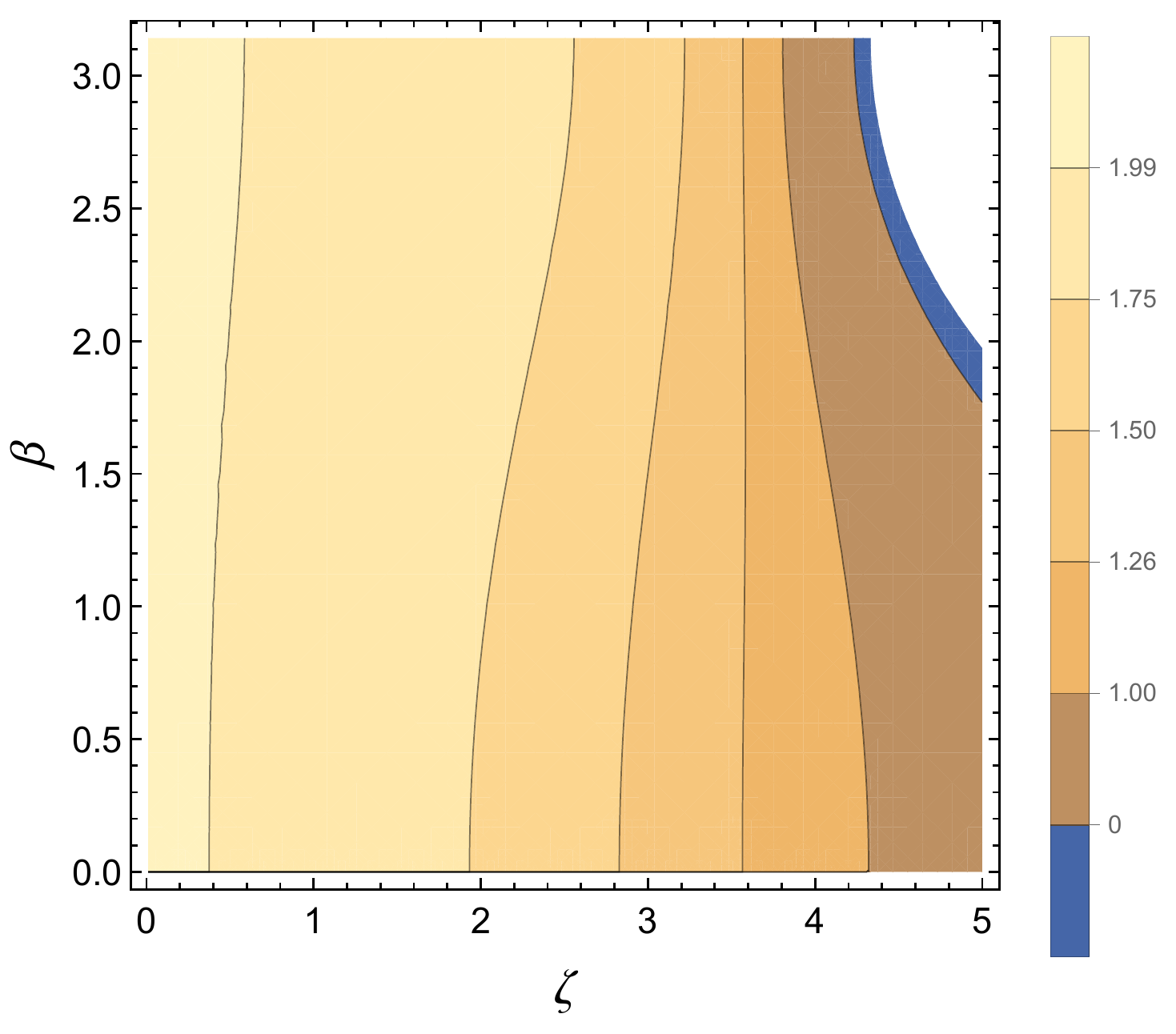}
\caption{The contour plot for the ratio $\Theta(\zeta,\beta)$.   We have the limit  $\lim_{\zeta\to 0}\Theta=2$  and almost vertical contour line around $\zeta=3.575$ (corresponding to 13Hz for ground-based detectors). }  
\label{fig:corrdif}
\end{figure}

\section{second generation detector network}\label{sec:4}

\begin{table}[t]
\caption{The latitudes, longitudes, and orientations of the five ground-based detectors in units of degree. The angle $\alpha$ is the orientation angle of the bisector of the two arms measured from the local east at each detector\footnote{\url{https://git.ligo.org/}}.}
\begin{ruledtabular}
\begin{tabular}{lccc}
	detector & latitude & longitude & $\alpha$\\ \hline \hline
	KAGRA(K)  & 36.41 & 137.31 &74.60\\
	LIGO-I(I) &  19.61 & 77.03 & 162.62\\
	LIGO-H(H) & 46.45 & -119.41 & 171.00\\
	LIGO-L(L) & 30.56 & -90.8 & 242.17\\
	Virgo(V) & 43.63 & 10.50 & 115.57\\
\end{tabular}
\end{ruledtabular}
\label{tab:position}
\end{table}

\begin{table*}[t]
\caption{(Upper right) The opening angle $\beta$ (in units of degree) of the detector pairs, measured from the center of the Earth. (Lower left) The values of $(\cos4\delta,\cos4\Delta,\sin 4\Delta)$.}
\begin{ruledtabular}
\begin{tabular}{lccccc}
	 & KAGRA & LIGO-I & LIGO-H & LIGO-L & Virgo\\ \hline \hline
	KAGRA & * &  54.89 & 72.37 & 99.27 & 86.52 \\
	LIGO-I &  (-0.41,0.63,0.78) & *  &  112.28 & 128.47  & 59.79 \\
	LIGO-H & (0.99,-0.34,0.94) & (0.75,0.47,-0.88) & * & 27.22 & 79.62 \\
	LIGO-L & (-1.00,0.19,-0.98) & (-0.80,-0.06,1.00) & (-1.00,-0.40,-0.91)& * & 76.76\\
	Virgo & (-0.60,0.87,0.50)& (-0.99,0.14,-0.99)& (-0.43,-0.80,-0.60)  & (-0.31,0.86,-0.50) & *\\
\end{tabular}
\end{ruledtabular}
\label{tab:relposition}
\end{table*}

\begin{table*}[t]
\caption{The expansion  coefficients $(C_{I_T},C_{I_V,}C_{I_S})$ (upper right) and  $(C_{W_T},C_{W_V})$ (lower left).}
\begin{ruledtabular}
\begin{tabular}{lccccc}
	 & KAGRA & LIGO-I & LIGO-H & LIGO-L & Virgo\\ \hline \hline
	KAGRA & * &  (-0.54,0.43,-1.22) & (0.48,0.15,1.06) & (-0.34,-0.06,-0.39) & (-1.1,0.64,-0.77) \\
	LIGO-I &  (-0.54,0.70) & *  &  (-0.80,0.40,0.05) & (0.11,-0.06,-0.05)  & (-0.29,-0.53,-1.20)\\
	LIGO-H & (-0.93,0.90) & (1.55,-0.57) & * & (-0.11,-1.62,-1.00) & (0.86,-1.02,0.24)\\
	LIGO-L & (1.48,-0.78) & (-2.04,0.42) & (0.28,-0.51)& * & (-0.97,0.77,-0.83)\\
	Virgo & (-0.63,0.45)& (0.77,-0.93)& (0.68,-0.57)  & (0.54,-0.48) & *\\
\end{tabular}
\end{ruledtabular}
\label{tab:asympcoef}
\end{table*}
From now on, we mainly discuss the ground-based detector networks composed by the following five second generation interferometers;  LIGO-Handford (H),   LIGO-India (I),  KAGRA (K),  LIGO-Livingston (L) and Virgo (V). We present their basic angular parameters in Table~\ref{tab:position}.

From these five interferometers, we can make $\rm _5C_2=10$ pairs and {introduce the abstract  index $u$ to represent these ten pairs} $\{{\rm HI, HK,\cdots, LV}\}$. Their relative geometrical parameters are presented in Table~\ref{tab:relposition}.  

\begin{figure}[t]
\centering
\includegraphics[keepaspectratio, scale=0.5]{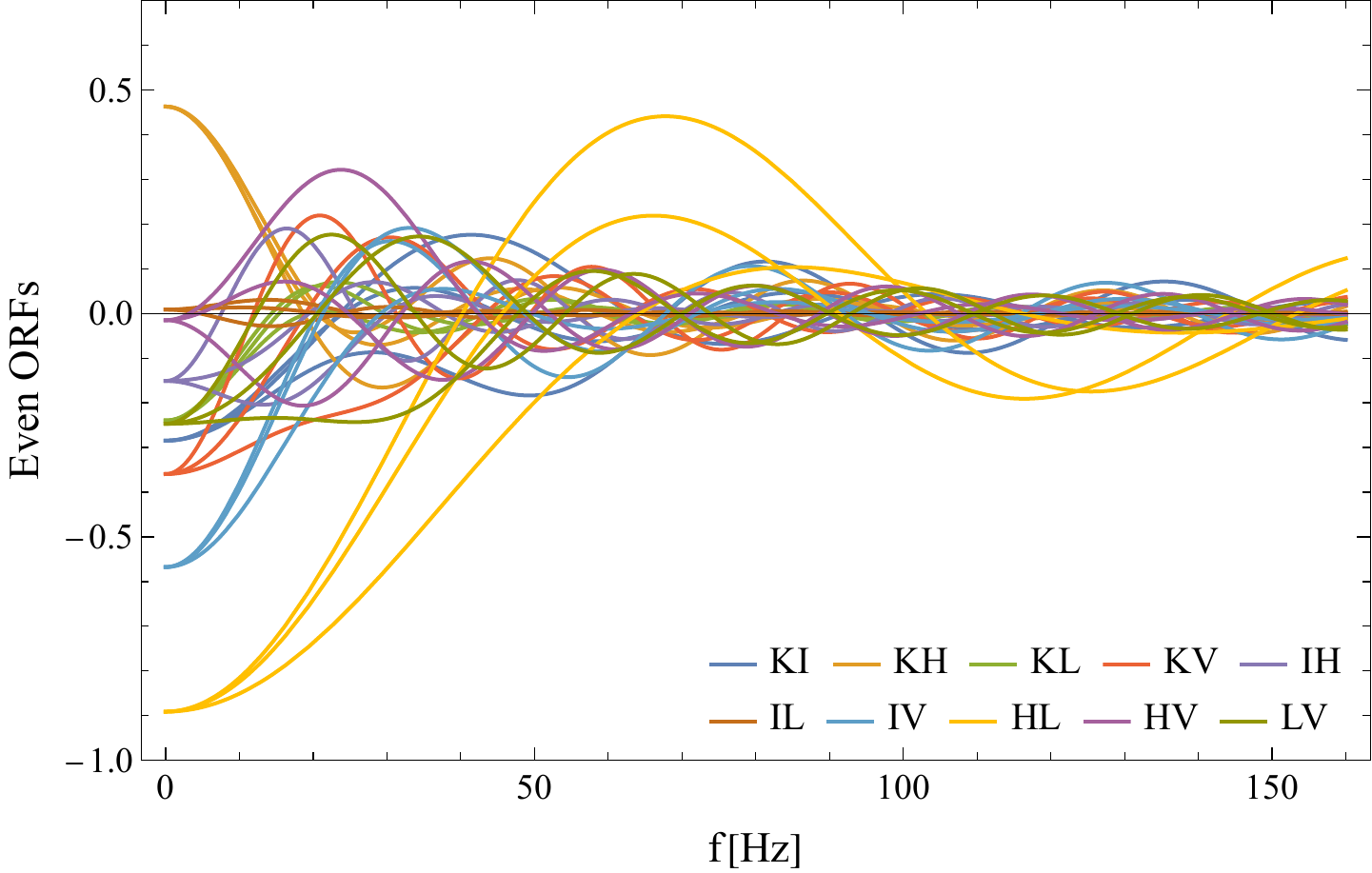}
\includegraphics[keepaspectratio, scale=0.5]{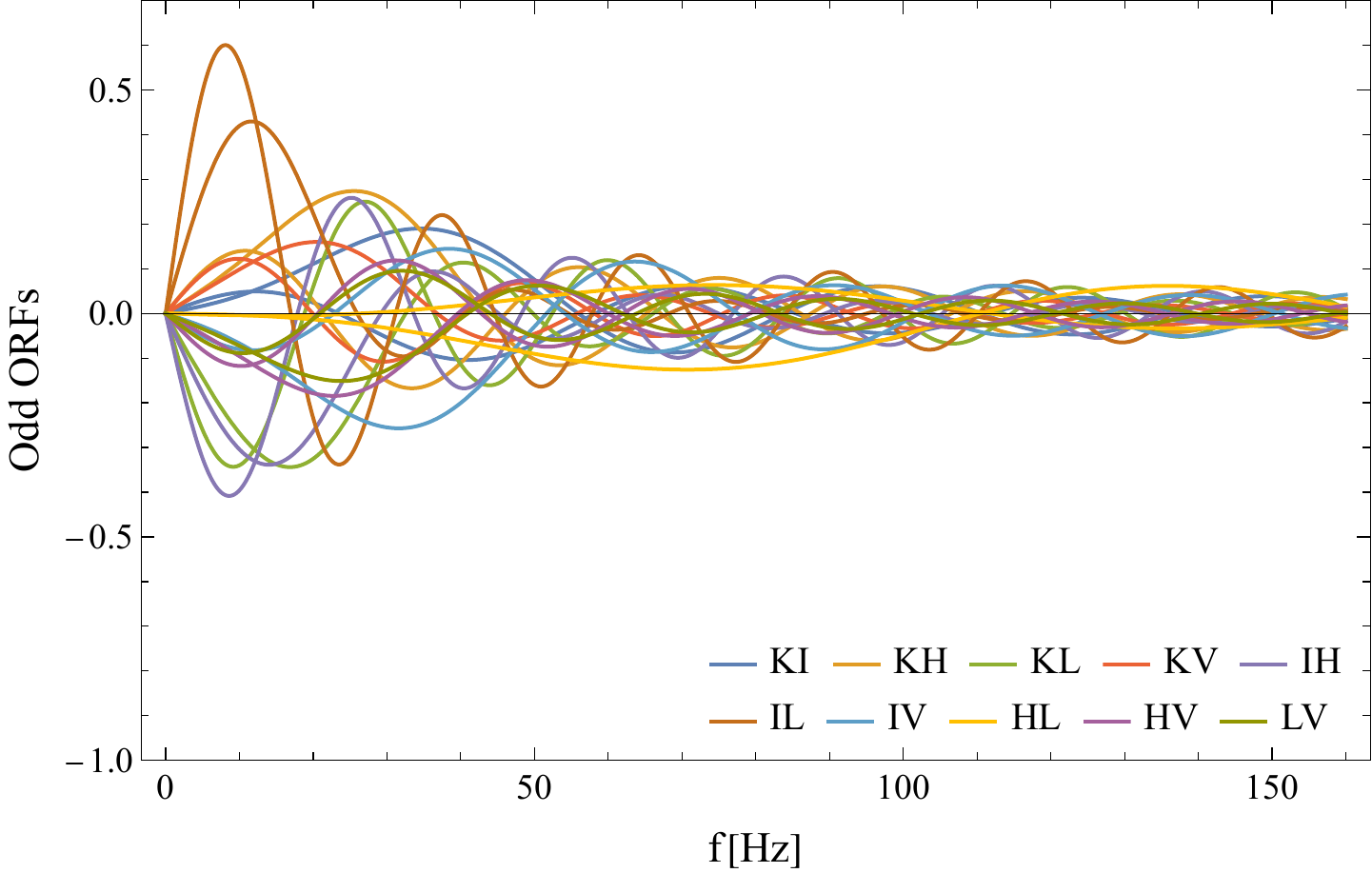}
\caption{All the 50 ORFs formed from the five interfeometers H, I,  K, L and V.  In the upper panel, the three curves starting  from $-0.9$  correspond to  the  HL pair.}
\label{fig:ORFall}
\end{figure}

Since each pair has the five ORFs $\gamma_u^Q(f)$ ($Q; I_T, I_V, I_S, W_T, W_V$), the total number of ORFs is 50.   In Fig. \ref{fig:ORFall}, we present all of them at a clip. Later, we will come to deal with the sums of their products such as $\sum_u\gamma_u^Q(f)\gamma_u^{Q'}(f)$, and the collective behaviours of these large number of ORFs would be important there. 

\begin{figure}[t]
\centering
\includegraphics[keepaspectratio, scale=0.4]{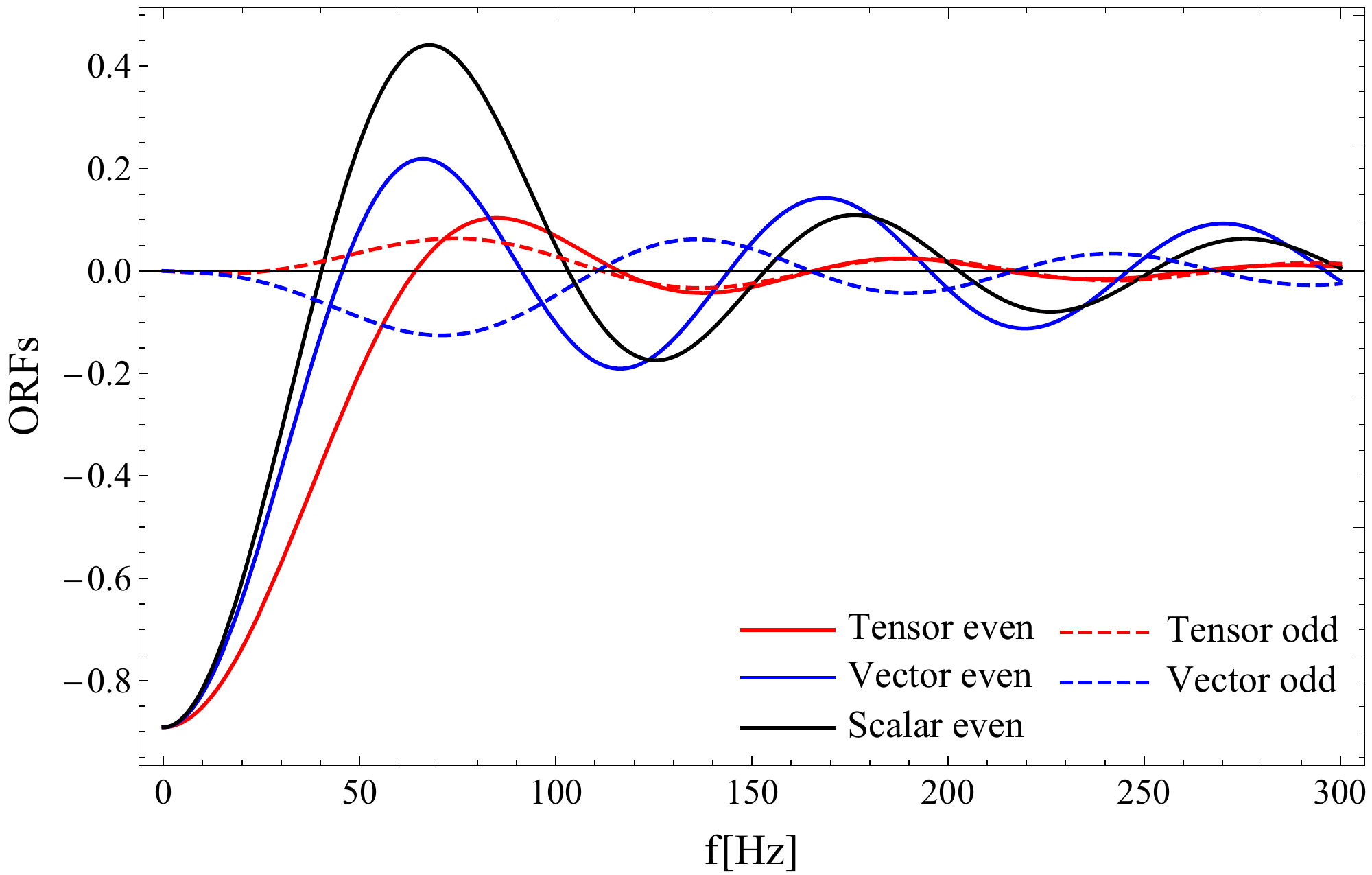}
\caption{All the five ORFs of the HL pair with $y=6.3 (f/100{\rm Hz})$. The  solid lines correspond to the even ORFs. The dashed lines show the odd ones. }
\label{fig:ORFHL}
\end{figure}

As  explained in Sec. II.E, at $f=0$, we have the degeneracies $\gamma_u^{I_T}=\gamma_u^{I_V}=\gamma_u^{I_S}$ and $\gamma_u^{W_T}=\gamma_u^{W_V}=~0$.   In Fig. \ref{fig:ORFall}, we can easily identify the three conspicuous curves starting from $\gamma_u^Q\simeq-0.9$ at $f=0$. These are  the even ORFs of the HL pair.   This pair is designed to have a large overlap with $\cos4\delta\simeq -1$.
 In Fig. \ref{fig:ORFHL}, its five ORFs are presented, showing loose oscillation patterns due to the small separation angle  $\beta$. The small  angle  $\beta$ also suppresses the amplitudes of the odd ORFs, in contrast to the even ones (see Sec.~\ref{sec:2D} and~\ref{sec:2E}).
 
 Meanwhile,  the HI and IL pairs have large separation angles $\beta$ and thus provide relatively large value 
 \begin{equation}
y\propto f\sin(\beta/2) \label{yf}
\end{equation}
  for a given frequency $f$. This will help us to use the higher order correction terms of  the variables $y$ (e.g., breaking the spectral degeneracy). 
  Together with the preferred relative orientation $|\sin4\Delta|\sim 1$, these pairs also have good sensitivities  to the odd parity  spectra $W_T$ and $W_V$.  
 
\begin{figure}[t]
\centering
\includegraphics[keepaspectratio, scale=0.55]{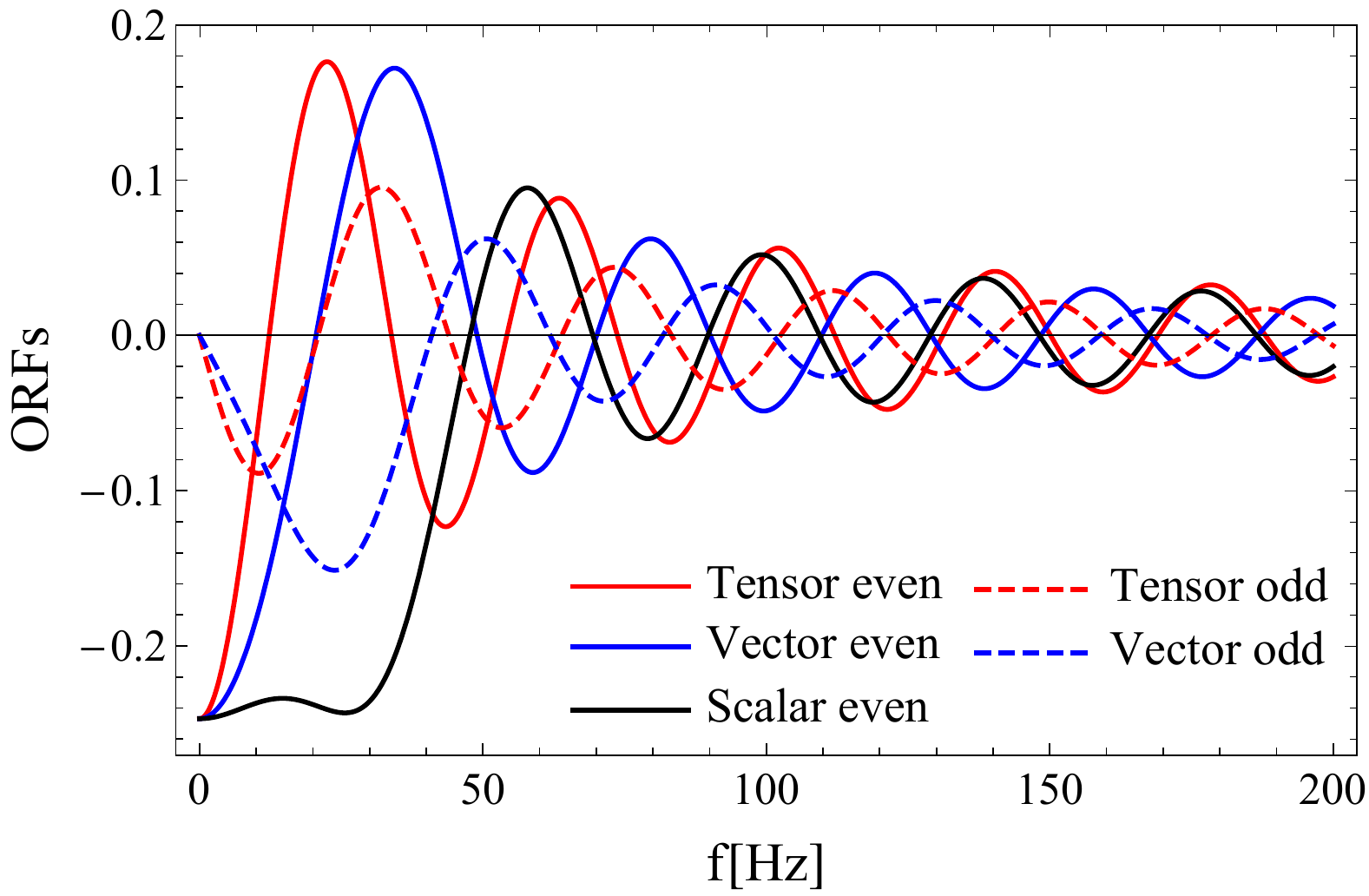}
\includegraphics[keepaspectratio, scale=0.55]{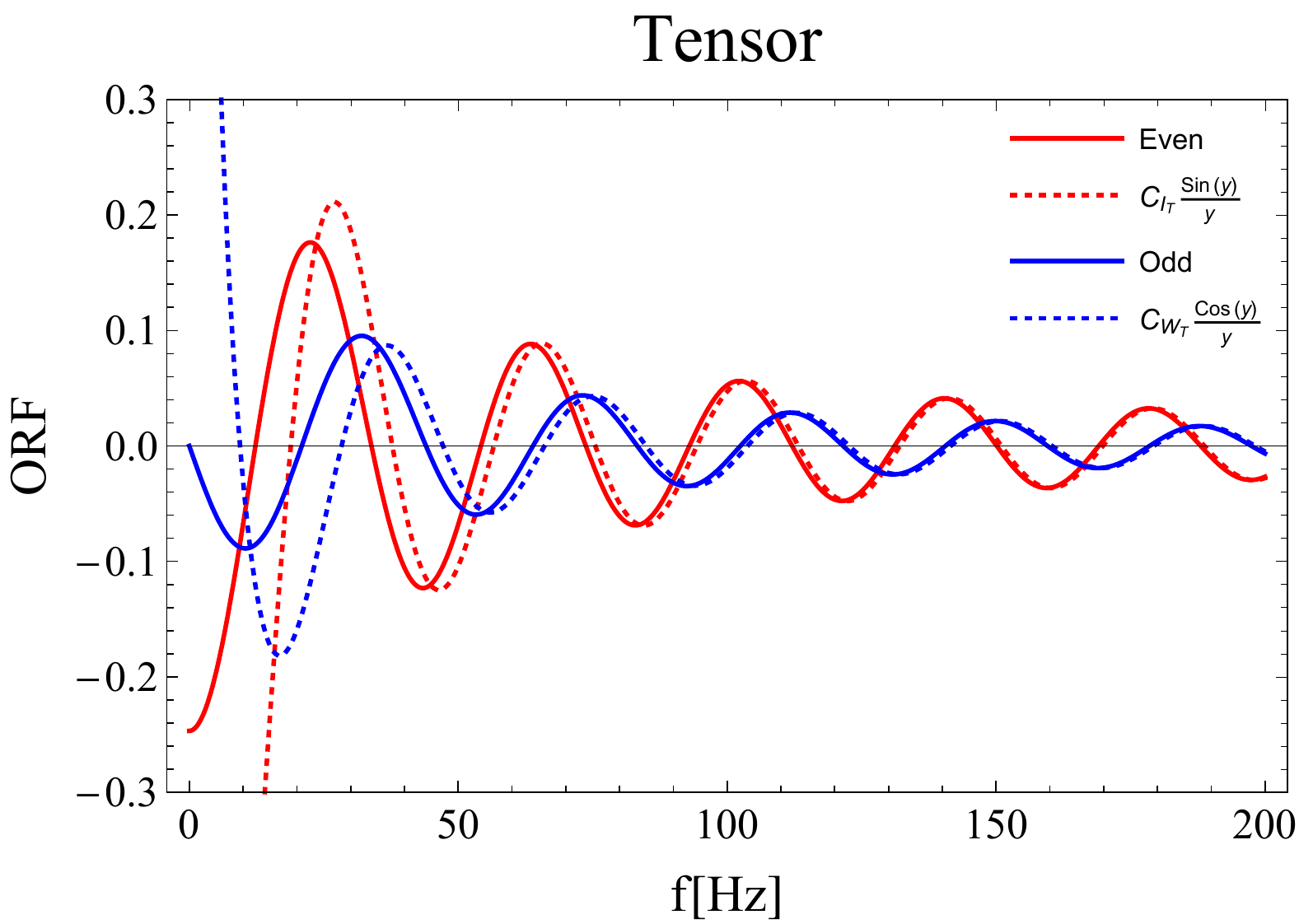}
\caption{The ORFs of the LV pair  with $y=16.65 (f/100{\rm Hz})$.  (Top) All the five ORFs. The solid lines correspond the even ORFs. The dashed lines are the odd ones. (Bottom)  The ORFs for the two tensor modes $I_T $ and $W_T$. The solid ones are the original expressions, and the dotted lines show their asymptotic profiles Eq. \eqref{eq:asymp} with the coefficients $(C_{I_T},C_{W_T})$ given in Table  \ref{tab:asympcoef}.}
\label{fig:ORFLV}
\end{figure}

  As examples of typical pairs, in Fig. \ref{fig:ORFLV}, we show the ORFs of the LV-pair. In the bottom panel, we compare the asymptotic profiles discussed in Sec. II.D.    At $f\gtrsim 80$Hz ($y\gtrsim 4\pi$), they show reasonable agreements with the original curves. Accordingly, in the upper panel, we can see the phase offset $\sim \pi/2$ between the odd and even ORFs  there. In  Table \ref{tab:asympcoef}, we present the asymptotic coefficients $C_Q$ for the ten pairs. 

\section{Correlation analysis with ground-based detectors}\label{sec:4A}

Up to this point, we only considered the response of  detectors to  stochastic backgrounds. In reality, the data streams of the detectors are contaminated by the detector noises. As we see  below, the correlation analysis is a powerful framework to coherently amplify the background signals relative to  the noises  \cite{Flanagan:1993ix,Allen:1997ad}. 

Under the existence of the detector noises, the outputs of two detectors $A$ and $B$ can be modeled as
\begin{align}
	s_A(f) &= h_A(f) + n_A(f)~, & s_B(f) &= h_B(f) + n_B(f)~.
\end{align}
Here, $h_{A,B}$ are the signals from stochastic backgrounds (see Eq. \eqref{eq:GWsig}) and $n_{A,B}$ are the detector noises. In this paper, we assume the noises $n_{A,B}$ to be stationary, Gaussian, and mutually independent. In addition, the signals are assumed to be much smaller than the noises, namely $|h_{A,B}| \ll |n_{A,B}|$ (the weak signal condition). Then the covariance of the detector noises is given by
\begin{align}\label{eq:noisespec}
	\braket{n_A (f)n_B^*(f')} = \frac{\delta_{AB}}{2}N_A(f)\delta(f-f')~,
\end{align}
where $N_A$ is the noise power spectrum.

As a preparation of  the correlation analysis, let us take the product of the two outputs of pairwise detectors ($u = AB$) as (again omitting the delta functions)
\begin{align}\label{re}
	\mu_{u}(f) \equiv \mathrm{Re}[s_A(f) s_B^*(f)]~.
\end{align}
Here we extracted the  real part. This is because,  we know the following relation
\beqa
\braket{s_A(f)s^*_B(f)} 
	&=& 
\braket{h_A(f)h_B(f)} + \braket{h_A(f)n_B(f)} \\
	& &+ \braket{n_A(f)h_B(f)} + \braket{n_A(f)n_B(f)}]\cr
	& =&\braket{h_A(f)h_B(f)} \\
	&=& C_u(f) \in {\rm Real}
\eeqa
for the expectation value (using the  statistical independence between $h_{A,B}$ and  $n_{A,B}$).  As we see shortly, this projection can reduce the associated noise level \cite{Seto:2006hf}. 

The variance can be calculated similarly. Under the weak signal condition ($|h_A(f)| \ll |n_A(f)|$),  we have
\begin{align}\label{eq:varianceab}
	\mathcal{N}_{u}(f) =& \braket{\mu_{u}^{2}} - \braket{\mu_{u}}^2 \sim \braket{\mu_{u}^{2}}\cr
	=& \frac{1}{4}\braket{(s_As_B^* + s_A^*s_B)(f)(s_As_B^* + s_A^*s_B)(f)}\cr
	\sim& \frac{1}{2}\braket{n_A(f)n_B^*(f) n_A^*(f) n_B(f)}\cr
	=& \frac{1}{8} N_A(f) N_B(f) ~
\end{align}
with  $\sqrt{\mathcal{N}_u(f)} \gg \lla \mu_u(f)\rra $.  Note that we have the additional factor $2^{-1}$ due to the real projection  \eqref{re}.

The basic idea of the correlation analysis is to coherently amplify the  background signal relative to the noise, by using   a large number of Fourier modes, after a long observational time. We now explain this by deriving  Eqs.  \eqref{eq:59} and \eqref{eq:varianceab}.

To deal with the frequency dependence, we first divide the Fourier modes  into $N$ 
 bins $(B_1,B_2,\cdots,B_\rho,\cdots ,B_N)$ characterized by the central frequencies $f_\rho$ and a fixed width $\delta f$ \cite{Seto:2006hf}. 
 We take $\delta f$ to be much smaller than $f_\rho$, such that involved quantities  ({\it e.g.} $I^{P}(f),W^P(f),$ and $\gamma^{I_P,W_P}(f)$) are nearly the same  in each bin.  Meanwhile, we also set the width $\delta f$ to be much larger than the frequency resolution $T_{\rm obs}^{-1}$ determined by the observation time $T_{\rm obs}$ ({\it i.e.} the number of the modes $T_{\rm obs}\delta  f\gg 1$ in each bin).

Now, we sum up the product $\mu_u$ in each bin as
\begin{align}
	\mu_{u}^{\rho} =& \sum_{f\in B_\rho}\mathrm{Re}[s_A(f)s_B(f)^* ]\cr
	 \simeq &\sum_{f\in B_\rho}\mathrm{Re}[h_A(f)h_B(f)^* +n_A(f)n_B(f)^* ]\label{eq:estsum}\\
	  \simeq &\braket{\mu_{u}^{\rho}}
	 +\sum_{f\in B_\rho}\mathrm{Re}[n_A(f)n_B(f)^* ]~.\label{eq:estsum2}
\end{align}
In Eq. \eqref{eq:estsum}, the first term comes  from the background and can be coherently amplified. On the other hand, the second term is due to  the noises and is not amplified because  of  its incoherence.

Let us calculate the expectation value and the variance of the compressed estimator $\mu_u^\rho$.   From Eqs. \eqref{eq:GWcorr} and   \eqref{eq:GWexp},  the expectation value $\braket{\mu_u^\rho}$ is given by
\begin{align}
	\braket{\mu_{u}^{\rho}} =& \sum_{f\in B_\rho} \mathrm{Re}[\braket{h_A(f)h_B(f)^* }] \cr
	\sim& \frac{8\pi}{5} T_{obs}\delta f \left(\sum_{P = T,V,S} \gamma^{I_P}_{u}(f_\rho) I_P(f_\rho) \right.\cr
	&\left. + \sum_{P=T,V} \gamma^{W_P}_{u}(f_c) W_P(f_c)\right)	\label{eq:59} ~.
\end{align}
The variance is given by the second term in Eq. \eqref{eq:estsum} as
\begin{align}\label{eq:varianceab}
	\mathcal{N}_{u}^{\rho} =& \braket{\mu_{u}^{\rho2}} - \braket{\mu_{u}^{\rho}}^2\cr
	\sim& \frac{1}{2}\sum_{f}\sum_{f'}\braket{n_A(f) n_A^*(f')} \braket{n_B(f) n_B^*(f')} \cr
	\sim& \frac{T_{obs} \delta f}{8} N_A(f_\rho) N_B(f_\rho) ~
\end{align}
with no noise correlation between different pairs (e.g., between HK and HL).
The last line is obtained by substitution of Eq. \eqref{eq:noisespec}. These expressions show that expectation value $\braket{\mu^\rho_{u}}$ is proportional to the number of the Fourier modes $T_{obs}\delta f$ but the variance $\sqrt{\mathcal{N}^\rho_u}$ is proportional to $\sqrt{T_{obs} \delta f}$. For $T_{obs}\delta f \gg 1$, the background signal is relatively amplified to the noise, as expected.

Combining Eqs. \eqref{eq:59} and \eqref{eq:varianceab}, we obtain the SNR of each bin as 
\begin{align}
	{\rm SNR}_{u}^{\rho 2} &= \frac{\braket{\mu_{u}^{\rho}}^2}{\mathcal{N}_{u}^{\rho}}\cr
	&\sim 2 \left(\frac{16 \pi}{5}\right)^2 \frac{T_{obs} \delta f}{N_A(f_\rho) N_B(f_\rho)} \left(\sum_{P = T,V,S} \gamma^{I_P}_{u}(f_\rho) I_P(f_\rho) \right.\cr
	&\left.+ \sum_{P=T,V} \gamma^{W_P}_{u}(f_\rho) W_P(f_\rho)\right)^2~.
\end{align}
Quadratically summing up all the frequency bin, we obtain total SNR for the detector pair $u$ as
\begin{align}\label{eq:SNRsingle}
	{\rm SNR}_{u}^2 & = \sum_{\rho} {\rm SNR}_{u}^{\rho2} \cr
	&= 2 T_{obs} \left(\frac{16 \pi}{5}\right)^2\cr
	& \times \int df \frac{\left(\sum_{P = T,V,S} \gamma^{P}_{u} I_P + \sum_{T,V} \gamma^{W_P}_{u} W_P\right)^2}{N_A(f) N_B(f)}~.
\end{align}

\begin{figure}[t]
\centering
\includegraphics[keepaspectratio, scale=0.42]{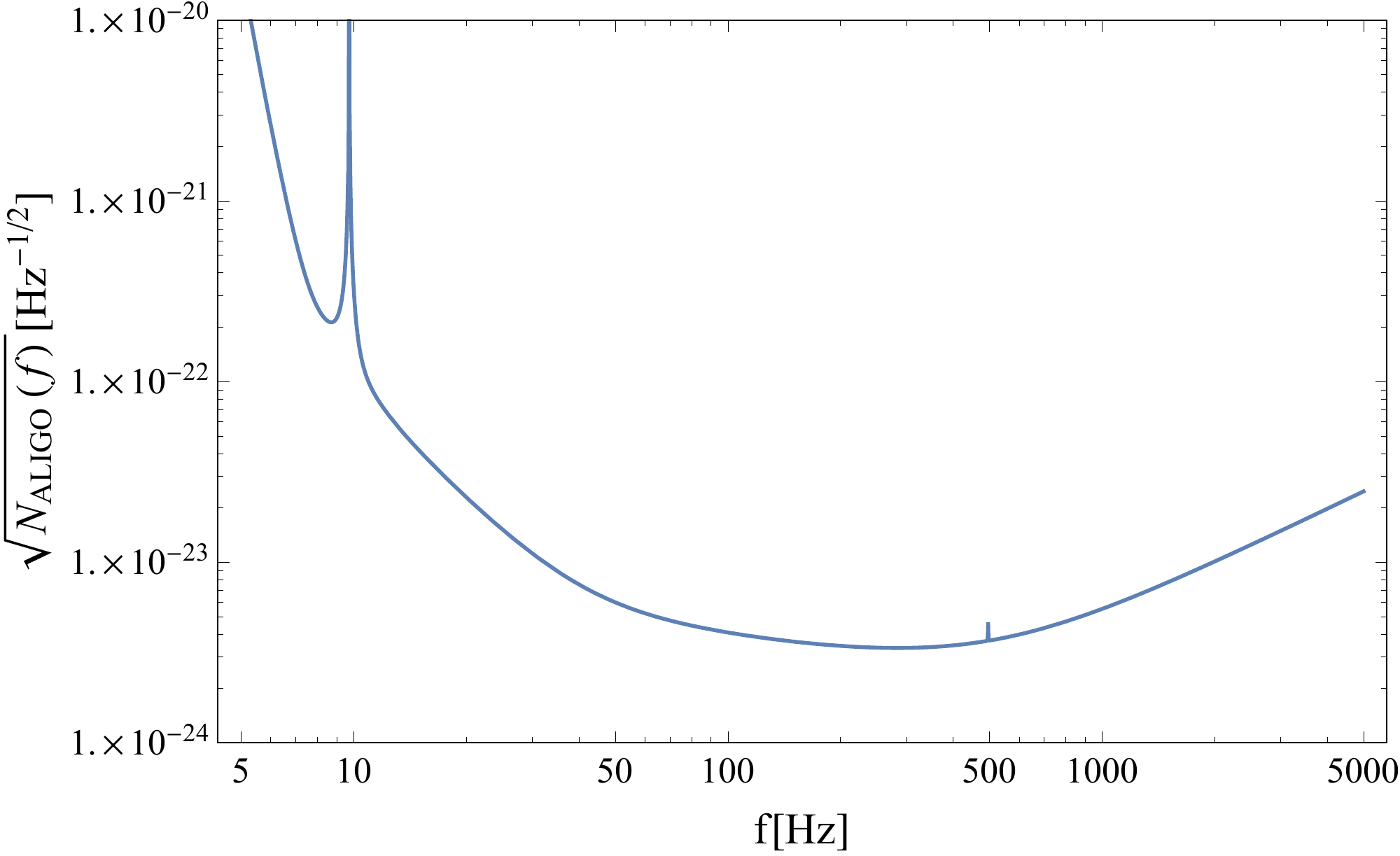}
\caption{Noise power spectrum of advanced LIGO, taken from \cite{aLIGOsensitivity}. {The spike around 9Hz is due to the resonance of the anti-vibration components.}}
\label{fig:noise}
\end{figure}

In this paper, we assume that  all detectors have the noise spectrum $N_{\rm AL}$ identical to  the design sensitivity of the advanced LIGO \cite{aLIGOsensitivity} (see Fig. \ref{fig:noise} for $N_{\rm AL}(f)$).
{Considering the current status of the LVK-network, this assumption looks unrealistic. However, it is virtually difficult  for a largely less sensitive detector to make an  effective contribution to the network, and we expect that our assumption will eventually become a reasonable approximation.}  
Note  that it is,  in principle, straightforward to taking into account the difference between detector noise spectra for the rest of this paper.  
 For simplicity,  unless otherwise stated, we also assume flat spectra $\Omega_{GW}^Q(f)\propto f^0$ for the injected backgrounds.

In the upper right panel of Table \ref{tab:SNR1}, we present SNR$_u$ for $\Omega_{GW}^{I_T}=10^{-8}$, setting other four spectra at zero. Similarly, in the lower left part, we show SNR$_u$ only with the non-vanishing compoent  $\Omega_{GW}^{W_T}=10^{-8}$ (ignoring the physical requirement $|\Omega_{GW}^{W_T}|\le \Omega_{GW}^{I_T}$).  In the ten detector pairs,  the HL pair has the best sensitivity to $I_T$, but the worst sensitivity to $W_T$. This is due to the small separation angle $\beta=27^\circ$ of the HL pair, as pointed out earlier in Sec. \ref{sec:2D}.   In  contrast, the IL pair has the worst sensitivity to $I_T$ but the best sensitivity to $W_T$ with the largest separation angle $\beta=128^\circ$.

\begin{table*}[t]\label{tab:SNR1}
\caption{The upper right corresponds to  ${\rm SNR}_{AB}$   for $\Omega^{I_T}_{GW}=10^{-8} $ setting other four spectra at zero.  The lower left  is only with 
$\Omega^{W_V}_{GW}=10^{-8} $.}
\begin{ruledtabular}
\begin{tabular}{lccccc}
	 & KAGRA & LIGO-I & LIGO-H & LIGO-L & Virgo\\ \hline \hline
	KAGRA & * & 2.16 & 2.42 & 1.38 & 4.47\\
	LIGO-I & 2.32& * & 2.79 & 0.34 & 3.27 \\
	LIGO-H & 3.67 & 5.07 & * & 15.4 &3.51 \\
	LIGO-L & 5.09 & 6.33 & 1.04 & * & 3.83\\
	VIRGO & 2.28 & 3.22& 2.56 & 2.07& *
\end{tabular}
\end{ruledtabular}
\end{table*}

For  a background purely made by $I_T$, we  have the network sensitivity
\begin{align}\label{eq:SNRref}
	{\rm SNR}_{I_T}^2 = 2 T_{obs} \left(\frac{16 \pi}{5}\right)^2 \int df \frac{\sum_{u} \left(\gamma^{I_T }_{u}\right)^2 I_T^2}{N_{\rm AL}^2(f)}~.
\end{align}
For the HIKLV network and a flat spectrum, we numerically have 
\begin{align}\label{eq:snmax}
{\rm SNR}_{I_T} = 19.0 \left(\frac{\Omega_{GW}^{I_T}}{10^{-8}}\right)\left(\frac{T_{obs}}{3 {\rm yr}}\right)^{1/2}(\equiv{\rm SNR}_0)~,
\end{align}
which gives the maximum sensitivity to $I_T$ achieved by the five detectors. In Eq. \eqref{eq:snmax},  we  introduced the notation ${\rm SNR}_0$  in order to use this result as a reference  value in our study below.

\section{Separation of the five components}\label{sec:4B}
As shown in Eq. (\ref{eq:59}), the expectation  value of a single  segment $\lla \mu_u^\rho(f)\rra$ is given as the linear combination of the five spectra $Q=\{I_T,I_V,I_S,W_T,W_V\}$.  For testing  alternative gravity theories, we would like to handle them separately. Such a method  has  been discussed in the literature ($I_T$ and $W_T$ in \cite{Seto:2006dz,Seto:2008sr}, and $I_T,I_V,$ and $I_S$ in \cite{Nishizawa:2009bf}). Its basic strategy is to take the appropriate linear combinations of the cross correlation signals $ \mu_u^\rho(f)$ and algebraically isolate the background spectra. 

To this end, we need at least 5 detector pairs. This can be satisfied by 4 or more detectors, which provide 6 or more pairs (not equal to 5). Thus the spectral decomposition is actually an overdetermined problem. 

{
Our first objective  in  this section is to present  a simple expression for the signal-to noise ratios ${\rm SNR}^\rho_{Q}$ after the algebraic spectral  decomposition. 
However,  as outlined in  Sec. VA,  under the orthodox approach,  we have a technical  difficulty at  deriving the simplified expression  ${\rm SNR}^\rho_{Q}$.   Thus, basically following the arguments  in  Ref. \cite{Seto:2008sr}, we  provide the  desired expression that is not proven in a precise mathematical sense.  
}

\subsection{Orthodox Approach}

As  an example,  let us consider the five detector network with 10 data set $\mu_u^\rho$ $(u=1,\cdots,10)$. Each segment  contains the five polarization  spectra as in Eq.~\eqref{eq:59}. Using the difference between the ORFs,  
we can isolate a specific spectrum $Q$ (e.g., $I_V$) by  algebraically cancelling other four spectra  (e.g., $I_T,I_S,W_T,W_V$). Then we obtain the six linear combinations of the original data    $\mu_u^\rho$.

In contrast to  the original ten data  $\mu_u^\rho$, the resultant six combinations have correlated detector noises. We  can newly generate six noise orthogonal combinations, as a standard eigenvalue decomposition for the $6\times 6$ noise matrix. Then,  quadratically adding the six  orthogonal elements, we obtain the network SNR for the target spectrum $Q$.   We can formally put  
 \begin{align}
({\rm  SNR}_Q^\rho)^2=2T_{obs}\delta f   \left(\frac{16 \pi}{5}\right)^2  \frac{Q(f)^2X_Q(f)}{N_{\rm AL}^2(f)}.  \label{x2}
\end{align}
Here the factor $X_Q(f)$  is  given by the 50 ORFs, and can be effectively regarded as the square of  a compiled ORF. 

Unfortunately, following the above line of argument, we could not analytically obtain the simplified symmetrical form for  the factor $X_Q(f)$  even with Mathematica.

\subsection{Alternative Approach}

In  Ref. \cite{Seto:2008sr}, a convenient construction scheme was deduced for the factor  $X_Q$, on the basis of the likelihood study for  the multiple  spectra (closely related to the  Fisher matrix analyses). Here we concisely provide their final expression (see   \cite{Seto:2008sr} for detail).

We first compose  a $5\times5$ matrix $F$ as 
\begin{align}\label{eq:covariance}
	F^{QQ'} \equiv  \sum_{u=1}^{n_p}\gamma^{Q}_u \gamma^{Q'}_u~.
\end{align}
Next, we take its inverse matrix 
\begin{equation}
\Sigma\equiv F^{-1}~.
\end{equation}

Then, we presume the following  relation for the factor $X_Q$
\beq
X_Q=\frac1{\Sigma^{QQ}(f)}. \label{simp}
\eeq
Below, we mention some circumstance evidences for its validity.

For decomposing only two spectra  (e.g., $I_T$ and $W_T$), we can analytically  confirmed that  this relation is  actually true for an arbitrary number  of detectors. For the five spectral decomposition with ten detector pairs, we numerically  generated the 50  ORFs  randomly in the range $[-1,1]$ and evaluated the both sides of  Eq. \eqref{simp} with Mathematica. We repeated this experiments for many times and confirmed equality within numerical accuracy. Note that, with Mathematica, we need much less computational resources  at numerical  evaluation than at corresponding symbolic processing.

We hereafter use relation  \eqref{simp} and put
 \begin{align}
({\rm  SNR}_Q^\rho)^2=\frac{\delta f}f  Z_Q(f) \left(\frac{\Omega_{GW}^Q(f)}{10^{-8}}\right)^2 \left( \frac{T_{obs}}{\rm 3yr} \right)       , 
\end{align}
where we defined (e.g., with Eqs. (\ref{e17}) and (\ref{x2}))  
\begin{equation}
Z_Q(f)\equiv 3.7\times 10^{-82} {X_Q}(f) \left(\frac{f}{\rm 1 Hz} \right)^{-5} \left(\frac{N_{AL}(f)}{\rm 1 Hz^{-1}} \right)^{-2}  \label{79}~.
\end{equation}
{Here we used $H_0 = 70{\rm km~s^{-1}~Mpc^{-1}}$.}
This function $Z_Q(f)$ shows the contribution of  background signals  from various  frequencies.

After the frequency integral, we obtain
 \begin{align}
({\rm  SNR}_Q)^2=\int_0^\infty \frac{df}f  Z_Q(f) \left(\frac{\Omega_{GW}^Q(f)}{10^{-8}}\right)^2 \left( \frac{T_{obs}}{\rm 3yr} \right)       .  \label{int}
\end{align}

\section{Statistical loss associated with   the mode separation}\label{sec:7}

We  now examine the matrices $F$ and $\Sigma$, in particular the role of their off-diagonal elements. 

\subsection{Reduction Factors}

For simplicity,  we first deal with the two component analysis with the spectra $I_T$ and $Q'$ $(Q' =I_V,I_S$ or $ W_T)$. The $2\times2$ matrix $F$ is given by
\begin{align}
	F=& \left(
	\begin{array}{cc}
	\sum_{u=1}^{n_p}\gamma_u^{I_T} \gamma_u^{I_T} & \sum_{u=1}^{n_p}\gamma_u^{I_T}\gamma_u^{Q'} \\
	 \sum_{u=1}^{n_p}\gamma_u^{I_T}\gamma_u^{Q'} & \sum_{u=1}^{n_p} \gamma_u^{Q'} \gamma_u^{Q'} 
	\end{array}
	\right)~,
\end{align}
and we have
\begin{align}
X_{I_T}=\frac1{\Sigma_{I_TI_T}}=(1-R_{I_TQ'}^2)  \sum_{u=1}^{n_p}\gamma_u^{I_T} \gamma_u^{I_T}~,\label{x3}\\
X_{Q'}=\frac1{\Sigma_{Q'Q'}}=(1-R_{I_TQ'}^2)  \sum_{u=1}^{n_p}\gamma_u^{Q'} \gamma_u^{Q'}~.
\end{align}
Here we  defined  the coefficient $R_{I_TQ'}$ by 
\begin{align}
	R_{I_TQ'} \equiv& \frac{\sum_{u=1}^{n_p} \gamma^{I_T}_u \gamma^{Q'}_u}{\sqrt{\sum_{u=1}^{n_t} \left(\gamma^{I_T}_u\right)^2}\sqrt{ \sum_{u=1}^{n_p}\left(\gamma^{Q'}_u\right)^2}}~.
\end{align}
From Cauchy-Schwartz  inequality, we have $|R_{I_TQ'}|\le 1$  with equality only for two parallel vectors $\{  \gamma^{I_T}_u   \}$    and $\{  \gamma^{Q'}_u  \}$.   The coefficient  $R_{I_TQ'}$  represents the correlation between the two spectra and reduces SNRs  after the spectral decomposition through the factor $(1-R_{I_TQ'}^2)$ (see Eqs. \eqref{x2} and \eqref{x3}).   This factor shows the statistical loss associated with the decomposition. 

So far,  we discussed two component decomposition. When  the number  $n_Q$ of the target spectral components  is  larger than two ($n_Q>2$), 
we can similarly define the reduction factor  $1-R_{Q_i}^2$ ($i=1,\cdots,n_Q$) by
\begin{align}
1-R_{Q_i}^2=\frac{X_{Q_i}(f)}{\sum_u \gamma_u^{Q_i}\gamma_u^{Q_i}}=\frac1{(F^{-1})^{Q_iQ_i}F^{Q_iQ_i}}~.
\end{align}
Note that we omitted the subscripts other than the component of interest for the notational simplicity.
If the vectors $\{ \gamma_u^{Q_i} \}$  ($i=1,\cdots,n_Q$)  are close to linearly dependent, the matrix $F$ becomes nearly singular, and we could have $|(F^{-1})^{Q_iQ_i}F^{Q_iQ_i}|\gg 1$, resulting in a large signal loss $1-R_{Q_i}^2 \ll 1$. 
In this relation, our two new findings in Sec. II could play interesting roles, as explained in the next subsection. 

\subsection{Numerical Results}
In Fig. 7, we show the reduction factor $(1-R_{I_T}^2)$ for the two component models $\{I_T,I_V\}$ (upper) and 
$\{I_T,I_S\}$ (lower).  As shown in Eq. (\ref{ge0}), we have the degeneracy $\lim_{f\to 0} \{\gamma_u^{I_T}\}=\{\gamma_u^{I_V}\}=\{\gamma_u^{I_S}\}$ and need the sub-leading correction $O(f^2)$ to decompose the  two spectra. We thus have a significant suppression $(1-R_{I_T}^2)\lesssim 0.1$ at $f\lesssim10$Hz.

\begin{figure}
\centering
\includegraphics[keepaspectratio, scale=0.6]{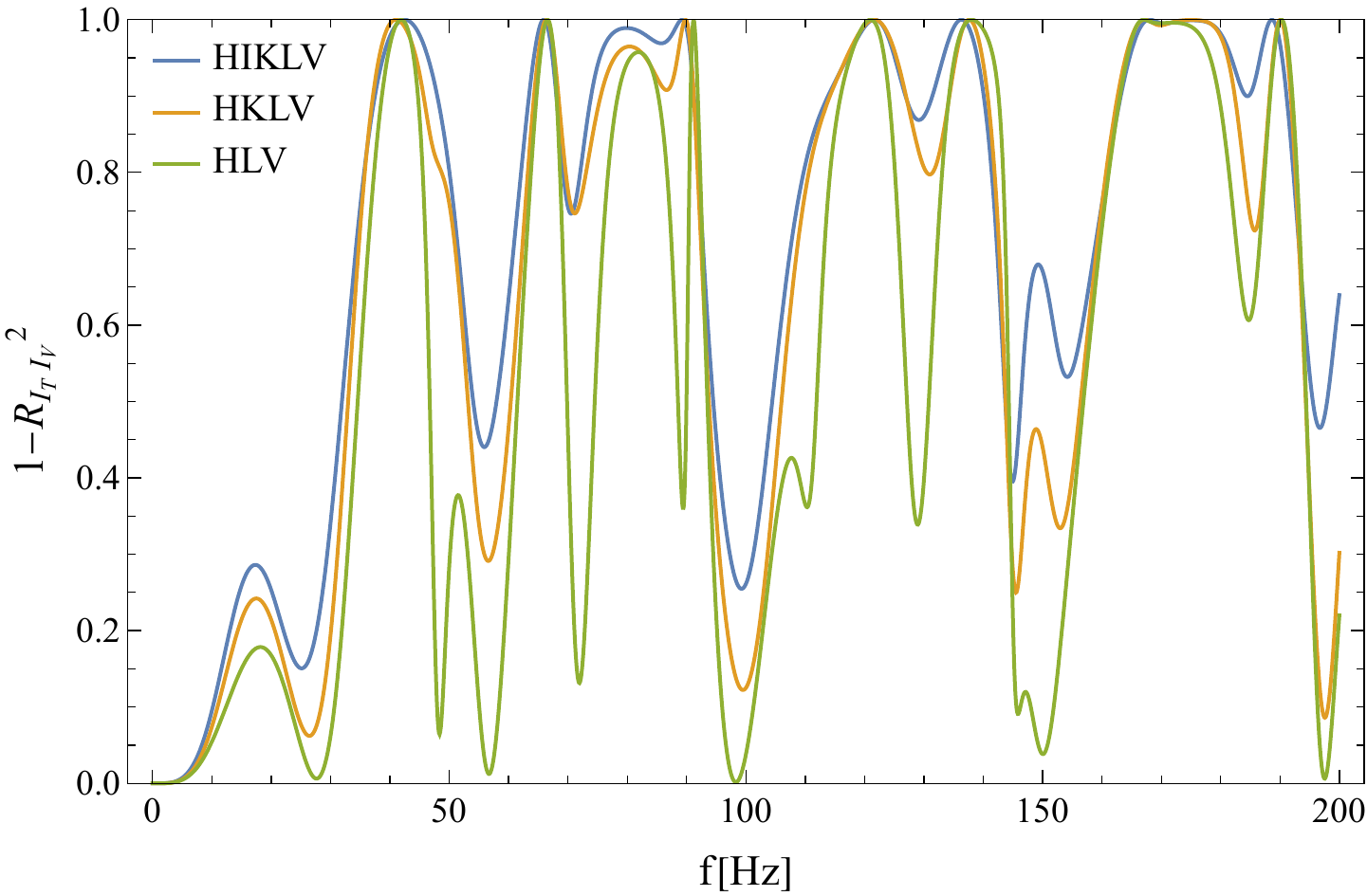}
\includegraphics[keepaspectratio, scale=0.6]{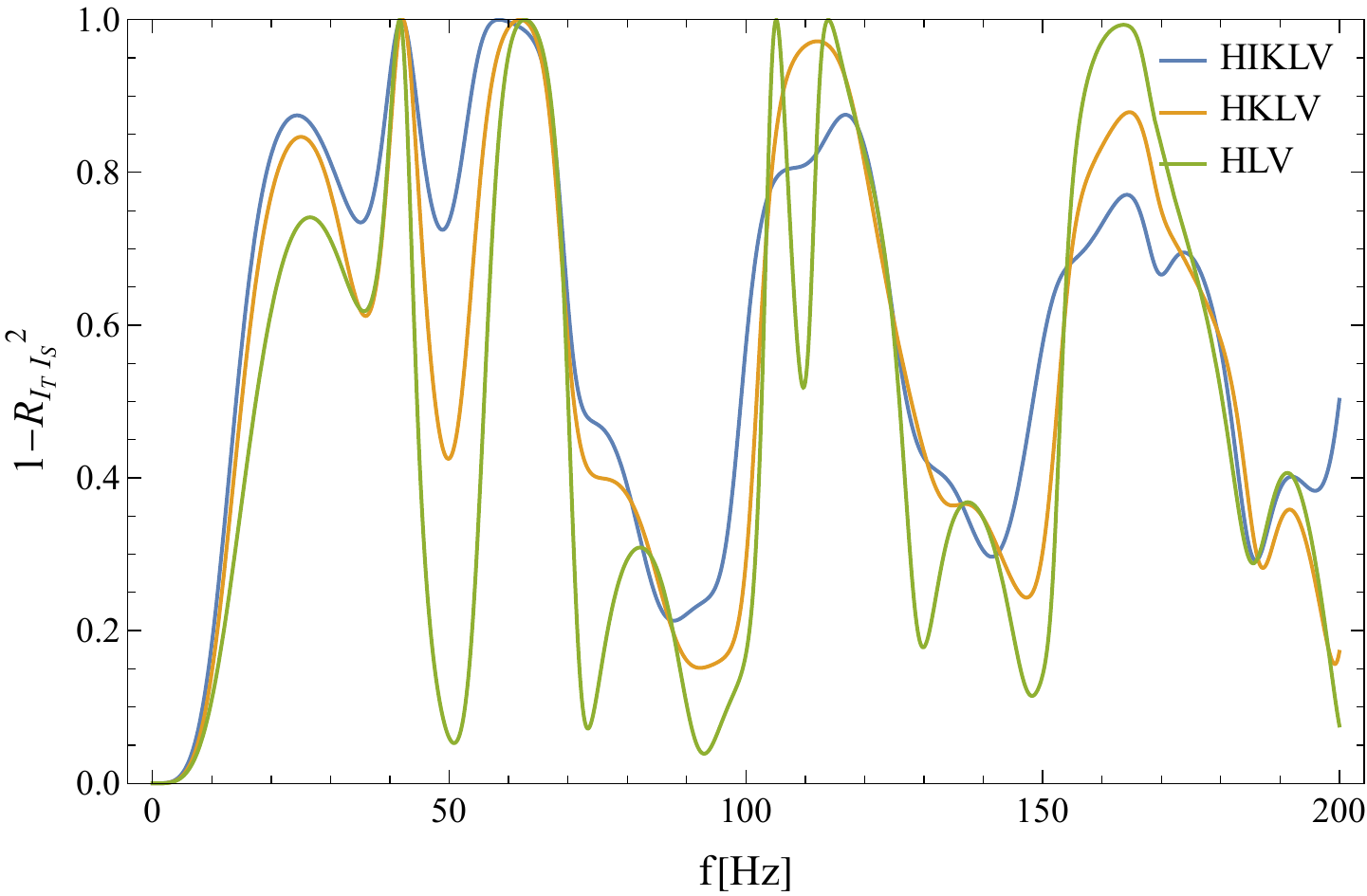}
\caption{The reduction factors $1-R_{I_T I_V}^2$ (upper) and $1-R_{I_T I_S}^2$ (lower)   respectively for  the two component analyses $\{I_T, I_V  \}$ and $\{I_T, I_S  \}$. 
}  
\label{fig:corrsame}
\end{figure}

\begin{figure}
\centering
\includegraphics[keepaspectratio, scale=0.6]{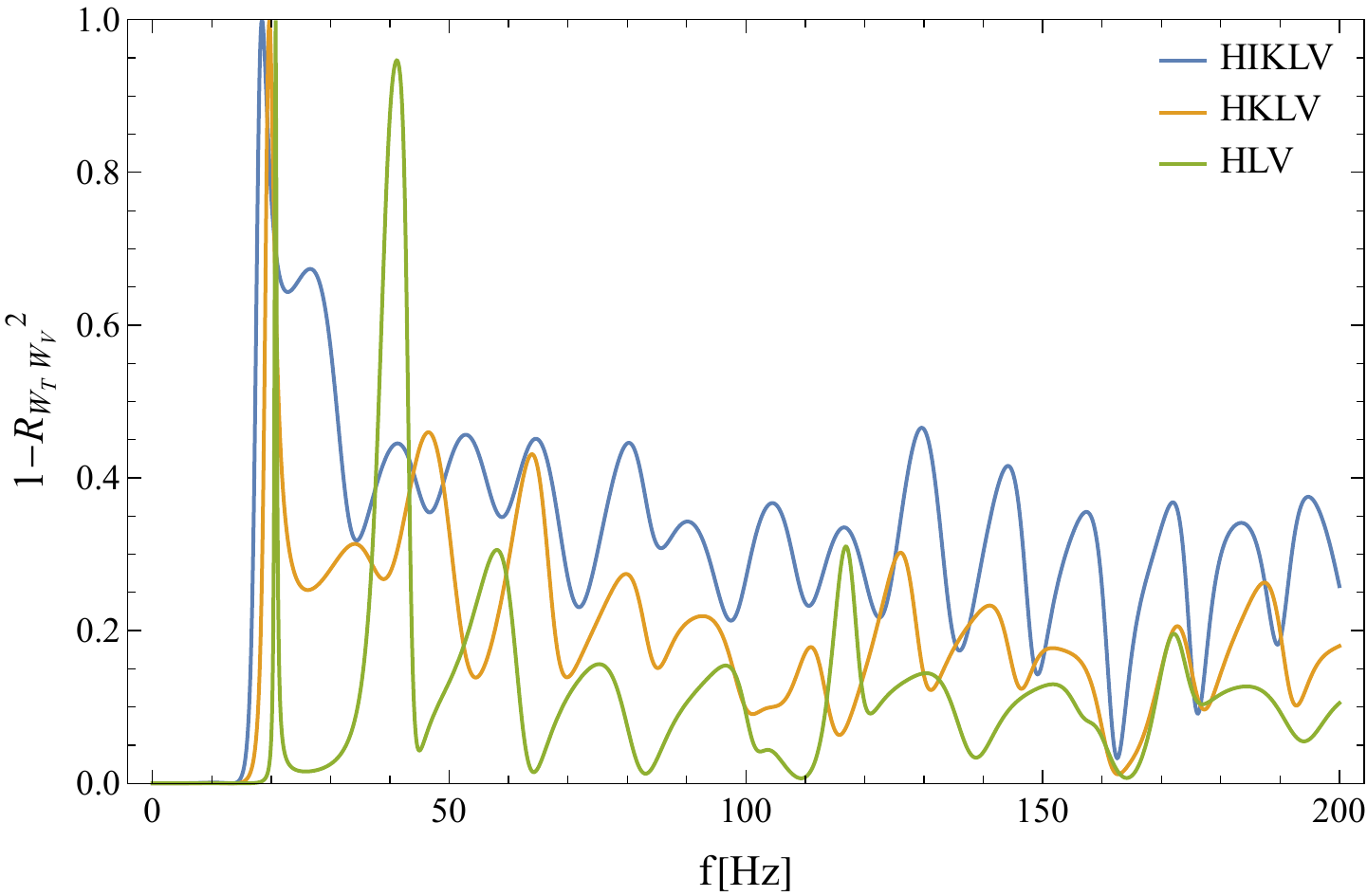}
\caption{The reduction factors $1-R_{W_T W_V}^2$ for the  hypothetical  two component analysis  $\{W_T, W_V  \}$. }  
\label{fig:corrdif}
\end{figure}

\begin{figure}
\centering
\includegraphics[keepaspectratio, scale=0.6]{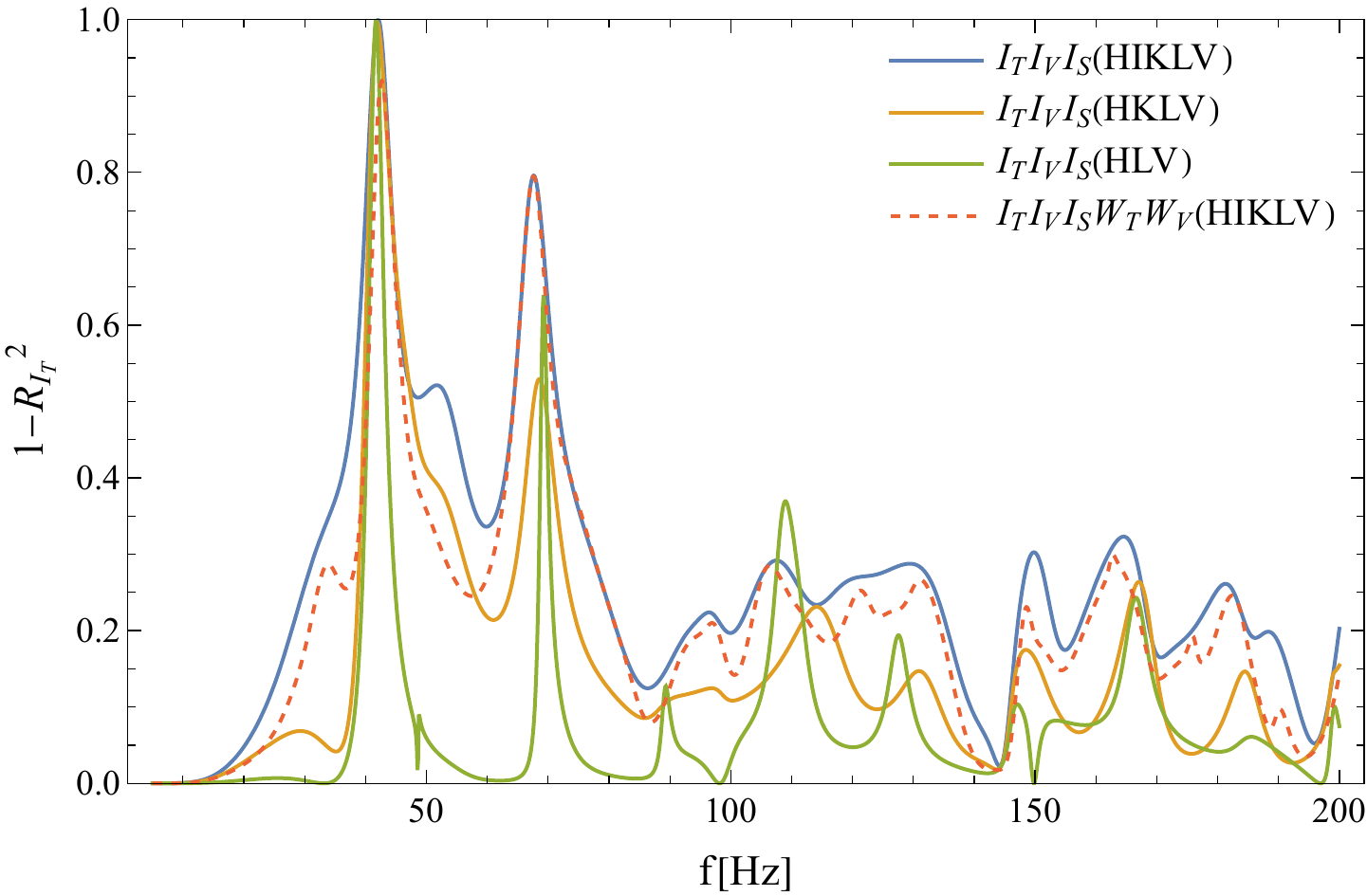}
\caption{The reduction factors $1-R_{T_T}^2$ for the three and five  component analyses. }  
\label{fig:corrdif}
\end{figure}

\begin{figure}
\centering
\includegraphics[keepaspectratio, scale=0.6]{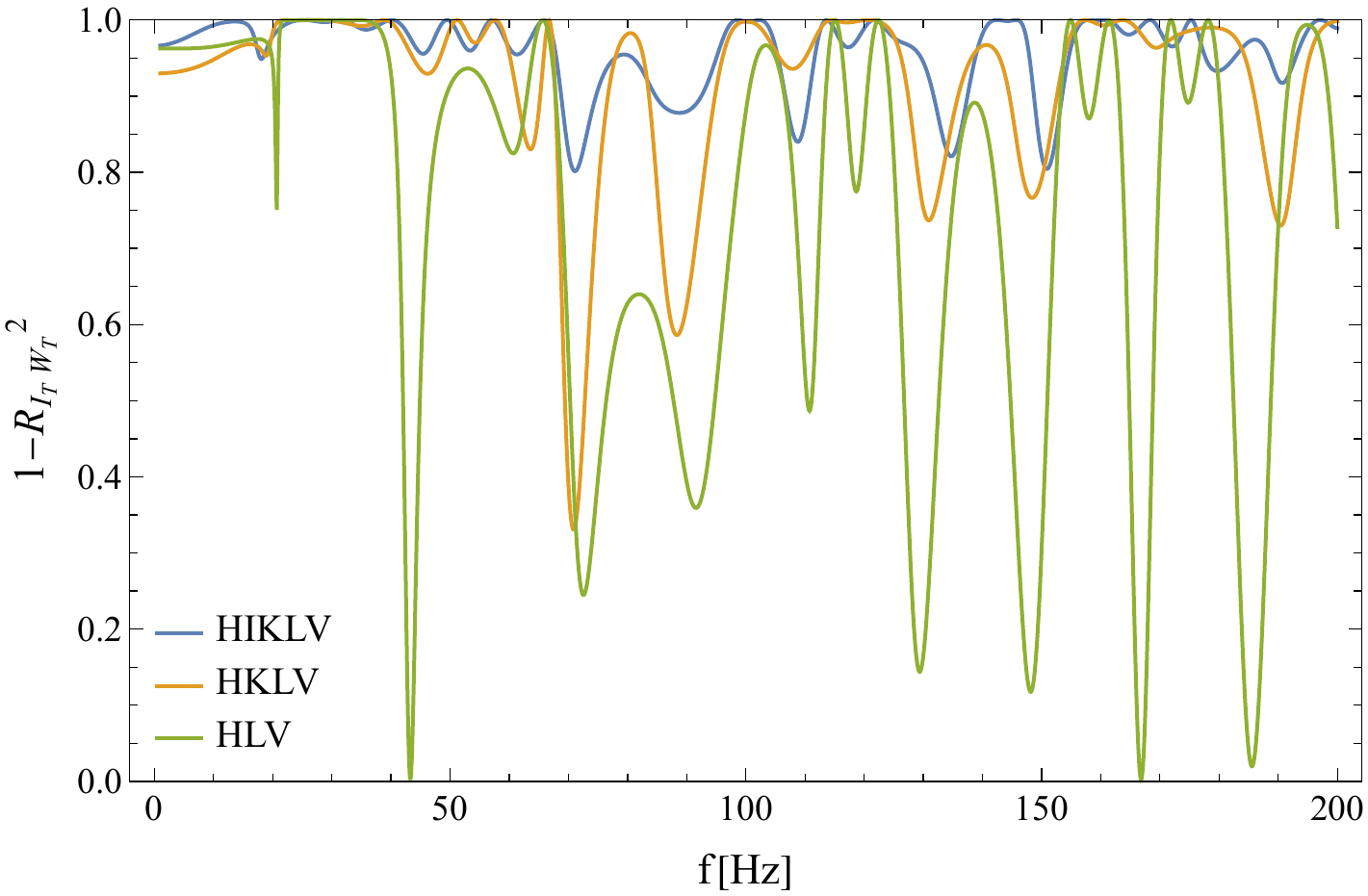}
\caption{The reduction factors $1-R_{I_T W_T}^2$ for the two component analysis  $\{I_T, W_T  \}$. }  
\label{fig:corrdif}
\end{figure}

In  Fig. 8,  we examined the hypothetical case for decomposing  the two odd spectra  $\{W_T,W_V\}$. Their ORFs are parallel at the low frequency limit and nearly parallel around the anathematic frequency 13Hz. We thus have the siginificant signal reduction below 13Hz,  as in Fig. 8.  

Next we move to examine the decomposition of  more than two spectra  $n_Q>2$. In Fig. 9, we show the reduction factor  $(1-R_{I_T}^2)$ at separating the three even spectra $\{I_T,I_V,I_S\}$.  In contrast to the two component cases in Fig. 7, the strong suppression  $1-R_{I_T}^2\lesssim 0.1$ continues up to $20$Hz.    This is because  the trinity  degeneracy \eqref{deg} works still at the sub-leading order $O(f^2)$. We thus need the higher corrections $O(f^4)$ to isolate the three spectra.   In fact, even for a detector network not tangential to a sphere, we still have $\det F=O(f^4)$ for the $3\times3$ matrix $F$ of  the three even spectra  $\{I_T,I_V,I_S\}$.

Note that the LIGO-India   plays a key role for the usage of the higher order terms  $O(f^4)$ (or more appropriately $O(y^4)$  for the perturbative expansion).    In Fig. 9, we can clearly see the resulting improvement  around 20-40Hz.   Here the mechanism around Eq. \eqref{yf} works efficiently, in particular, with the HI and IL pairs. 

In  Fig. 10, we present the result for the two tensorial spectra $\{I_T,W_T\}$. Since their ORFs $\{\gamma_u^{I_T}\}$ and $\{\gamma_u^{W_T}\}$   are generally not parallel, the reduction is not  significant.  
If we use the HIKLV pair, the reduction factor is no less than 0.8. 

\section{Signal to noise ratio}\label{sec:8}

\subsection{Results for the HIKLV Network}

Now we discuss the signal-to-noise ratios ${\rm SNR}_Q$ after the spectral decomposition  and the associated frequency    profiles $Z_{Q}(f)$   defined  in Eq. (\ref{79}).   We start with the results for the HIKLV network and flat spectra  $\Omega_{GW}^Q= {\rm const}$.    

 In  Fig. 11, we show the profile  $Z_{I_T}(f)$ for $I_T$.    The sharp dip around 10Hz is caused by the noise spike  in Fig. 5.   The uppermost blue line shows the result for the simplest case only with $I_T$ (no reduction factor). Its peak is around 25Hz with the integrated value ${\rm SNR}_{I_T}$ (see Eq.~\eqref{eq:snmax})
\begin{equation}
{\rm SNR}_0=19.0 \left(\frac{\Omega_{GW}^{I_T}}{10^{-8}}\right)\left(\frac{T_{obs}}{3 {\rm yr}}\right)^{1/2}.
\end{equation}
We use this expression to normalize  the signals ${\rm SNR}_Q$ in different settings, as presented in Tables V and VI. 
In Fig.  11, the  four lines other than the blue one show  the profiles  $Z_{I_T}(f)$ after decomposing multiple spectra. Their fractional differences  from the blue lines represent the corresponding reduction factor $(1-R_{I_T}^2)$. 

 At the decomposition of $I_T$ and $W_T$ (dashed orange line in Fig. 12), the statistical loss is inconspicuous with the total value ${\rm SNR}_{I_T}/{\rm SNR}_0=0.99$ (see  Table V).   
However,  we need to pay a significant cost to isolate the three even spectra $I_T$,  $I_V$ and $I_S$.   The total signal decreases  down to  ${\rm SNR}_{I_T}/{\rm SNR}_0=0.47$ and the peak  of  the profile $Z_{I_T}(f)$ moves up to $\sim 40$Hz.  

In Figure 12 and Table VI, we show the results for the odd tensor spectrum $W_T$. Similarly to Fig. 11,  we can isolate it from $I_T$ with almost no loss of  the integrated  signal ${\rm SNR}_{W_T}$ (see  Table VI).  
When we  separate $W_T$ and $W_V$, the anathematic frequency  13Hz clearly appears, as shown by the green and red lines,  and the function $Z_{W_T}(f)$ is significantly suppressed below $\sim20$Hz. In contrast to   $Z_{I_T}(f)$, the peak of the profile 
 $Z_{W_T}(f)$  stays around 25Hz. 

\begin{figure}
\centering
\includegraphics[keepaspectratio, scale=0.6]{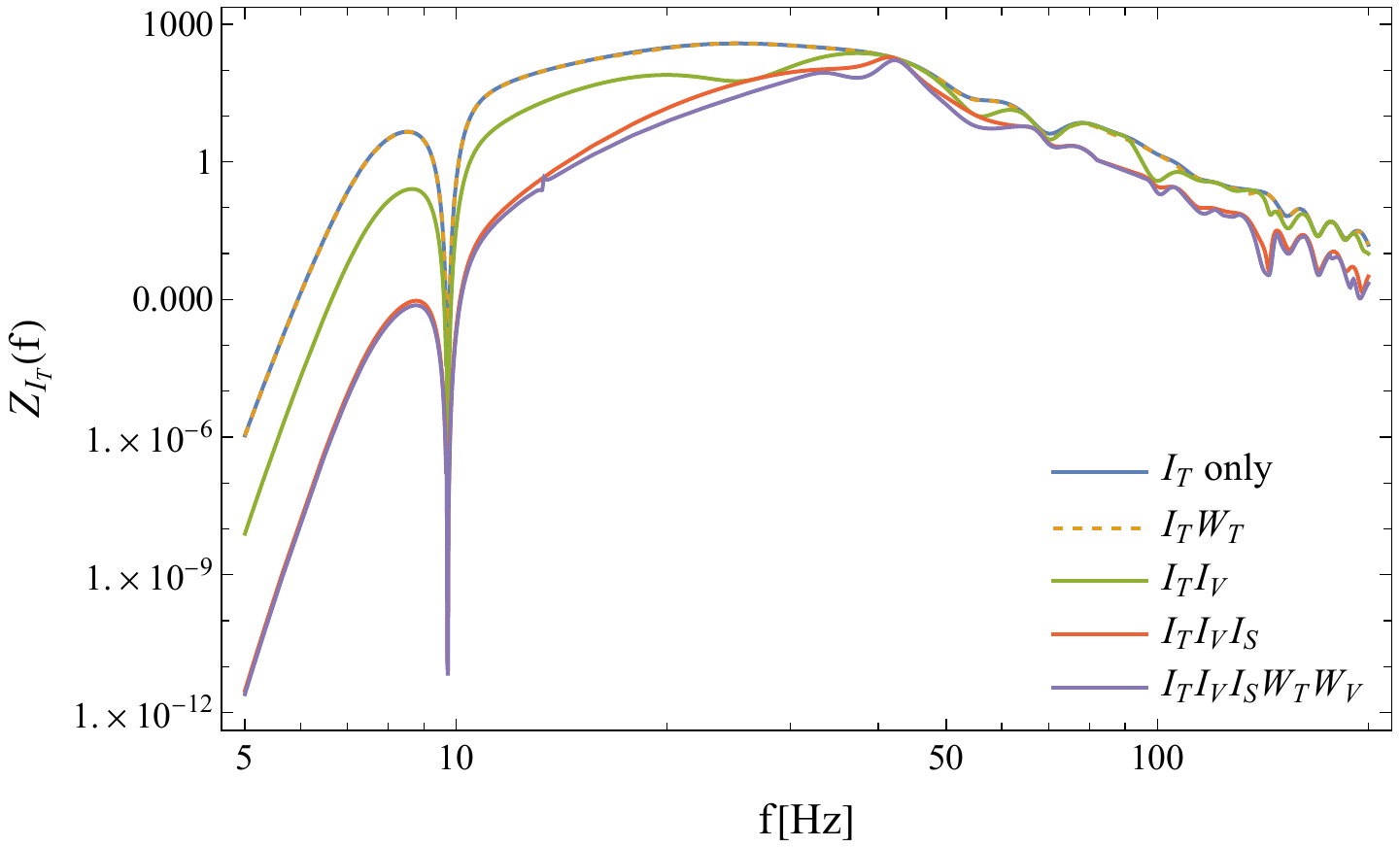}
\caption{ The factor $Z_{I_T}(f)$ showing the signal strength defined in Eq.  (\ref{79}) for the HIKLV network.   The blue and orange curves are nearly  overlapped. The ratio between the blue and other curves corresponds to the reduction factor $1-R_{I_T}^2$ due to the signal correlation. The sharp dip around 9Hz is due to the noise spectrum $N_{AL}(f)$.   }  
\label{fig:corrsame}
\end{figure}

\begin{figure}
\centering
\includegraphics[keepaspectratio, scale=0.6]{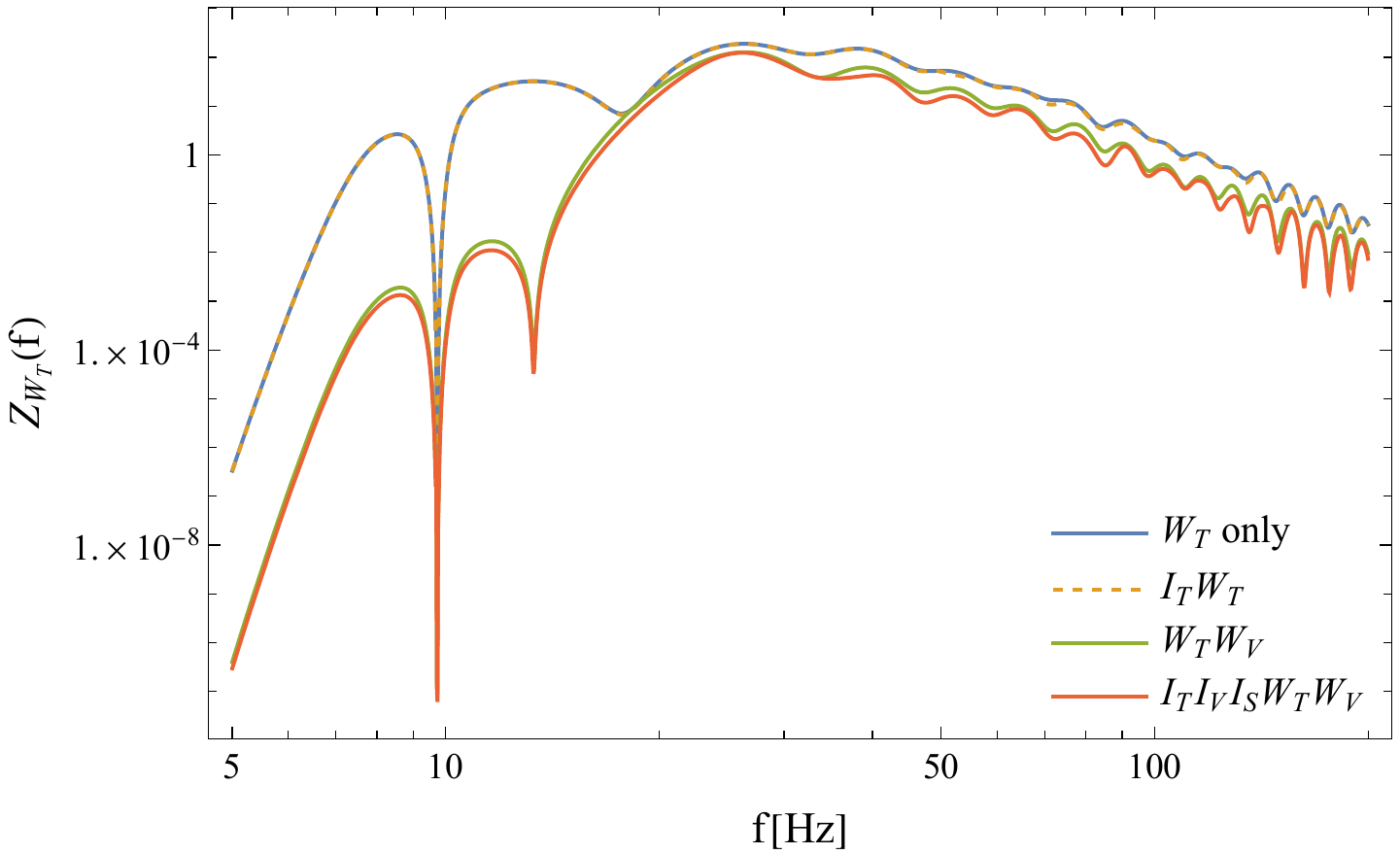}
\caption{The factor $Z_{W_T}(f)$ showing the signal strength defined in Eq.  (\ref{79}) for the HIKLV network.  The green and the red curves have the sharp dips around 13Hz.    }  
\label{fig:corrsame}
\end{figure}

\begin{figure}
\centering
\includegraphics[keepaspectratio, scale=0.60]{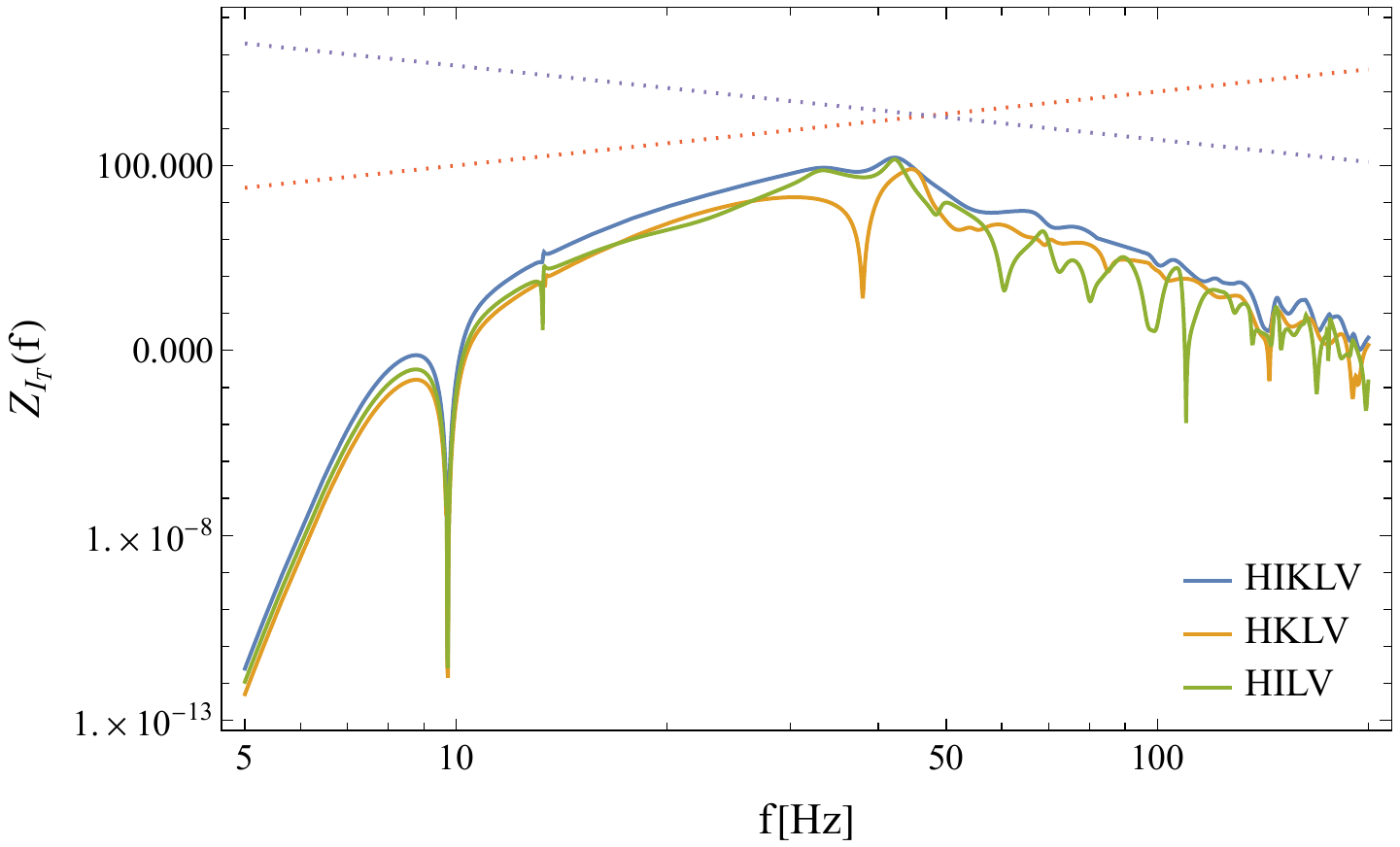}
\includegraphics[keepaspectratio, scale=0.60]{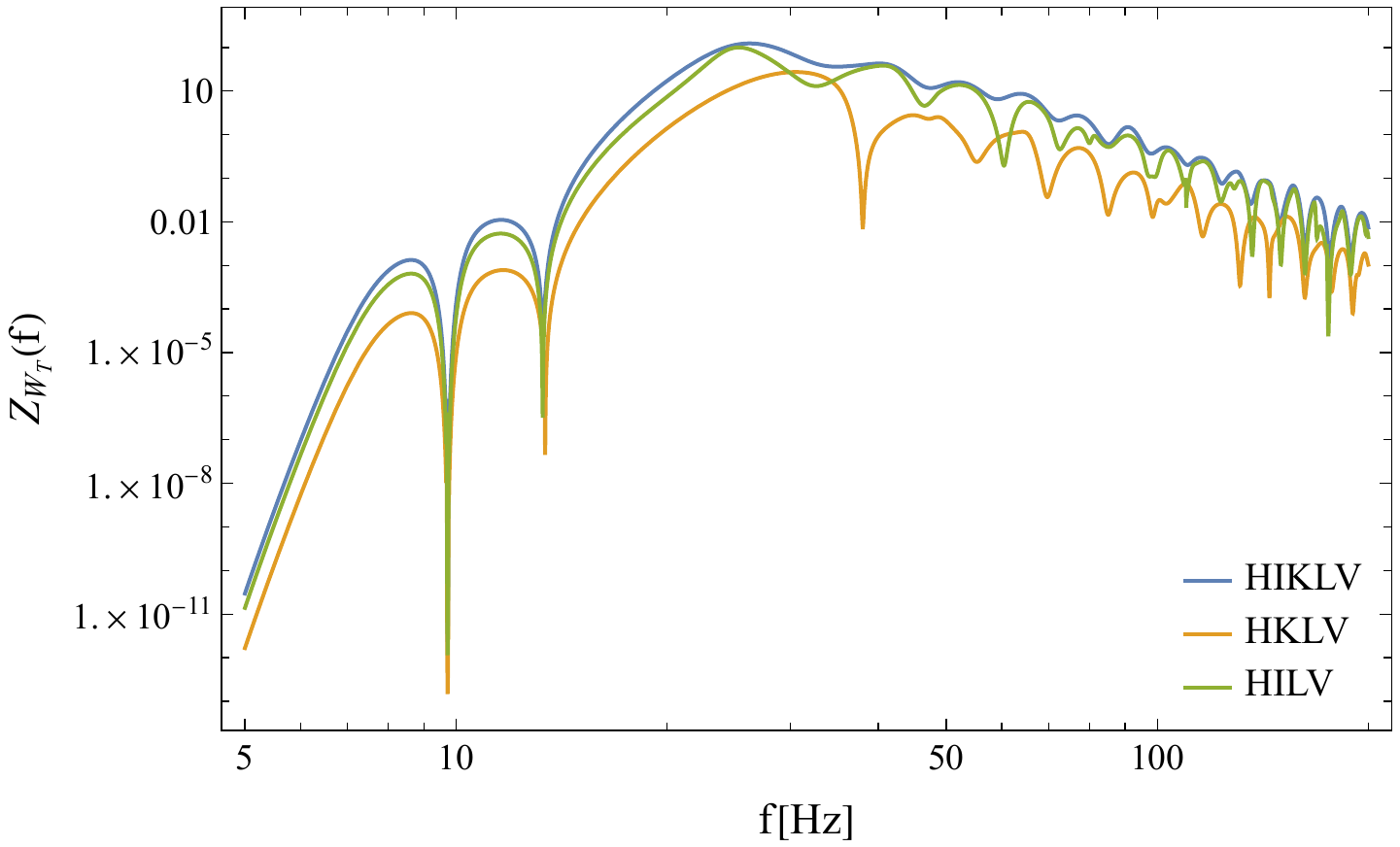}
\caption{The top and bottom panels respectively show $Z_{I_T}(f)$ and $Z_{I_T}(f)$  for three networks. The dotted lines in the upper panel are proportional to $f^2$  and $f^{-2}$.  }  
\label{fig:corrsame}
\end{figure}

\subsection{LIGO-India}

Next we  discuss  the impacts of adding LIGO-India  to the detector network. As shown in Table V,  for the single spectral search  $I_T$, LIGO-India increases the total signal $SNR_{I_T}$ by only 3\%. However, together with KAGRA, it makes a notable contribution to improve the sensitivity to the odd spectra $W_T$ (see Table VI). In addition, as explained earlier,  LIGO-India also helps us  to use the higher order terms $O(f^4)$ for decomposing the three even spectra.   For the  five spectral  search,   we can double both ${\rm SNR}_{I_T}$ and ${\rm SNR}_{W_T}$  by adding the LIGO-India detector. 

\subsection{Power-law Models}
So  far, we have assumed that the background has flat spectra $\Omega_{GW}^Q= {\rm const}$. Now, we briefly discuss a power-law form $\Omega_{GW}^Q\propto f^\alpha$  in the frequency regime in interest.

The integrated signal ${\rm SNR}_Q$ in Eq. (\ref{int})  has the dominant contribution around the frequency where the function $Z_Q(f)$ is tangential to a curve $f^{-2\alpha}$. As deduced from  Fig. 13, for the five spectral decompositions with the HIKLV network,  the tangential frequencies are 40Hz for  $I_T$ and 25Hz for $W_T$, as long as the index $\alpha$ is in the range $[-1,1]$.   Therefore, the total signals are very roughly given as 
\begin{eqnarray}
{\rm SNR}_{I_T}\sim 19\times 0.4 \left(\frac{\Omega_{GW}^{I_T}({\rm 40Hz})}{10^{-8}}\right)\left(\frac{T_{obs}}{3 {\rm yr}}\right)^{1/2}\\
{\rm SNR}_{W_T}\sim 19\times 0.39 \left(\frac{\Omega_{GW}^{W_T}({\rm 25Hz})}{10^{-8}}\right)\left(\frac{T_{obs}}{3 {\rm yr}}\right)^{1/2}.
\end{eqnarray}

\begin{table}[t]
\caption{Ratio ${\rm SNR}_{I_T}/{\rm SNR}_0$ after the spectral isolation.  All five components of LHV is missing since LHV has only three independent detector pairs. We assumed a flat spectrum $\Omega_{GW}^{I_T}
={\rm const}$. 
}
\begin{ruledtabular}
\begin{tabular}{lccc}
	background components & KILHV & KLHV & LHV\\ \hline \hline
	 $I_{T}$ only & 1 & 0.97 & 0.91\\
	$I_{T},W_{T}$ & 0.99 & 0.96 & 0.91\\
	$I_{T},I_{V}$ & 0.63 & 0.57 & 0.47\\
	$I_{T},I_{S}$ & 0.89 & 0.82 & 0.74\\
	$I_{T},I_{V},I_{S}$ & 0.47 & 0.33 & 0.22\\
	All five & 0.40 & 0.20 & *\\
\end{tabular}
\end{ruledtabular}
\label{tab:5}
\end{table}

\begin{table}[t]
\caption{Ratio ${\rm SNR}_{W_T}/{\rm SNR}_0$ after the spectral isolation.  We assume flat spectra and omit the factor $\Omega_{GW}^{W_T}/\Omega_{GW}^{I_T}$  for simplicity. }
\begin{ruledtabular}
\begin{tabular}{lccc}
	background components & KILHV & KLHV & LHV\\ \hline \hline
	 $W_{T}$ only & 0.62 & 0.40 & 0.18\\
	$W_{T},I_{T}$ & 0.62 & 0.39 & 0.18\\
	$W_{T},W_{V}$ & 0.43 & 0.20 & 0.06\\
	All five & 0.39 & 0.17 & *\\
\end{tabular}
\end{ruledtabular}
\label{tab:5}
\end{table}

\section{Summary}\label{sec:9}

In this paper, we studied the prospects for the polarizational study of isotropic stochastic gravitational wave backgrounds by correlating second generation detectors. 
 In the long-wave approximation,  the backgrounds are generally characterized by the five spectra $I_{T,V,S}$ and $W_{T,V}$. The modes other than $I_T$ can appear in modified theories of gravity. 

For correlation analysis, the ORFs play key roles. In this paper, we newly identified two simple relations behind them. The first one is the trinity degeneracy (\ref{deg}) between the three even ORFs at the sub-leading order $O(f^2)$. The second one is the degeneracy between the two odd  ORFs around the specific frequency 13Hz. 

For each detector pair, the correlation product is given as a linear combination of the five spectra. 
To closely examine theories of gravitation,  we desire to separate the five spectra clearly. 
We thus examined their algebraic decomposition using the difference between the involved  ORFs. Here we generally need to handle an over-determined problem. 
 By extending an analytic framework in the literature, we derived the formal expression (\ref{int}) for the optimal SNRs after the spectral decomposition. 
 
  Then, assuming an identical noise curve for the five detectors and flat background spectra,  we discussed the statistical loss of sensitivities accompanied by the decomposition.   This loss is closely related to the off-diagonal elements of the matrix $F^{QQ'}\propto \sum_{i=1}^{n_p} \gamma_u^Q \gamma_u^{Q'}$. 
  
  In this context, our two findings are quite useful for following the singular behaviors at the decomposition. 
  On the one hand, when simultaneously dealing with the three even spectra,  due to the higher order degeneracy of their  ORFs, we have a large signal reduction below 20Hz,  unlike the two spectral decomposition (such as $I_T-I_V$ and $I_T-I_S$). On the other hand, it is very difficult to separate the two odd spectra below $\sim 20$Hz, including the anathematic frequency 13Hz.   Given the structure of the covariance matrix $F$, these limitations will also appear in the likelihood or Fisher matrix analyses. 
  
We also discussed the advantage of adding the LIGO-India detector to the ground-based detector network. As shown in Tables V and VI, it can largely increase the sensitivities to the odd spectra and will also help us to decompose multiple spectra. Here the HI and LI pairs are particularly useful with the large separation angles $\beta$.  

In this paper, we have mainly considered the second generation ground-based detectors. However, our method is general enough to be straightforwardly applied to the third generation ground-based detectors (such as ET~\cite{Punturo:2010zz} and CE~\cite{Reitze:2019iox}, see also ~\cite{Amalberti:2021kzh}) and partially to space borne detectors (LISA~\cite{Audley:2017drz}, TAIJI~\cite{Hu:2017mde}, and TianQin~\cite{Luo:2015ght}). The former will cover a lower frequency regime than that of the second generation ones and will be more severely affected by the limitations associated with our two findings.

\acknowledgments

 We would like to thank M. Ando and  S. Bose  for valuable comments.   This work is supported by JSPS Kakenhi Grant-in-Aid
for Scientific Research (Nos. 17H06358 and 19K03870). HO is supported by Grant-in-Aid for JSPS Fellows JP22J14159.
 
 \appendix

\section{optimal SNR for the ground-based detectors}

Assuming the flat spectrum of the background and using Eq. \eqref{int}, we can evaluate  the SNR for each spectra after the decomposition.  As a reference, we provide numerical results for the five spectral components.    For the HKLV-network, we obtain
\begin{align}
	{\rm SNR}_{I_T} &= 3.94\left(\frac{\Omega_{GW}^{I_T}}{10^{-8}}\right)\left(\frac{T_{obs}}{3 {\rm yr}}\right)^{1/2}\\
	{\rm SNR}_{I_V} &= 2.75 \left(\frac{\Omega_{GW}^{I_V}}{10^{-8}}\right)\left(\frac{T_{obs}}{3 {\rm yr}}\right)^{1/2}\\
	{\rm SNR}_{I_S} &= 6.81 \left(\frac{\Omega_{GW}^{I_S}}{10^{-8}}\right)\left(\frac{T_{obs}}{3 {\rm yr}}\right)^{1/2}\\
	{\rm SNR}_{W_T} &= 3.14 \left(\frac{\Omega_{GW}^{W_T}}{10^{-8}}\right)\left(\frac{T_{obs}}{3 {\rm yr}}\right)^{1/2}\\
	{\rm SNR}_{W_V} &= 4.07 \left(\frac{\Omega_{GW}^{W_V}}{10^{-8}}\right)\left(\frac{T_{obs}}{3 {\rm yr}}\right)^{1/2}.
\end{align}
For the HIKLV-network, we have
\begin{align}
	{\rm SNR}_{I_T} &= 7.53 \left(\frac{\Omega_{GW}^{I_T}}{10^{-8}}\right)\left(\frac{T_{obs}}{3 {\rm yr}}\right)^{1/2}\\
	{\rm SNR}_{I_V} &= 6.14 \left(\frac{\Omega_{GW}^{I_V}}{10^{-8}}\right)\left(\frac{T_{obs}}{3 {\rm yr}}\right)^{1/2}\\
	{\rm SNR}_{I_S} &= 9.74\left(\frac{\Omega_{GW}^{I_S}}{10^{-8}}\right)\left(\frac{T_{obs}}{3 {\rm yr}}\right)^{1/2}\\
	{\rm SNR}_{W_T} &= 7.50 \left(\frac{\Omega_{GW}^{W_T}}{10^{-8}}\right)\left(\frac{T_{obs}}{3 {\rm yr}}\right)^{1/2}\\
	{\rm SNR}_{W_V} &= 8.27 \left(\frac{\Omega_{GW}^{W_V}}{10^{-8}}\right)\left(\frac{T_{obs}}{3 {\rm yr}}\right)^{1/2}.
\end{align}

\bibliography{ref}

\end{document}